\documentclass[aps,prb,reprint,showpacs,twocolumn,superscriptaddress,floatfix]{revtex4-1}
\usepackage{amssymb}
\usepackage{amsmath}
\usepackage{graphicx}
\usepackage{extarrows}
\usepackage[toc,page,title,titletoc,header]{appendix} 
\usepackage[usenames,dvipsnames]{xcolor}

\begin{document}

\title{Dynamical Coulomb blockade theory of plasmon-mediated light emission from a tunnel junction}

\author{F. Xu}

\affiliation{Fachbereich Physik, Universit\"at Konstanz, D-78457 Konstanz, Germany}

\author{C. Holmqvist}

\affiliation{Department of Physics and Electrical Engineering, Linnaeus University, 39182 Kalmar, Sweden}

\author{G. Rastelli}

\affiliation{Fachbereich Physik, Universit\"at Konstanz, D-78457 Konstanz, Germany}

\author{W. Belzig}

\affiliation{Fachbereich Physik, Universit\"at Konstanz, D-78457 Konstanz, Germany}

\email[]{Wolfgang.Belzig@uni.kn}

\date{\today}

\begin{abstract}
Inelastic tunneling of electrons can generate the emission of photons with energies intuitively limited by the applied bias voltage. However, experiments 
indicate that more complex processes involving the interaction of electrons with plasmon polaritons lead to photon emission with overbias energies. We recently proposed a model of this observation in Phys. Rev. Lett. \textbf{113}, 066801 (2014), in analogy to the dynamical Coulomb blockade, originally developed for treating the electromagnetic environment in mesoscopic circuits. This model describes the correlated tunneling of two electrons interacting with a local plasmon-polariton mode, represented by a resonant circuit, and shows that the overbias emission is due to the non-Gaussian fluctuations.  Here we extend our model to study the overbias emission at finite temperature. We find that the thermal smearing strongly masks the overbias emission. Hence, the detection of the correlated tunneling processes requires temperatures $k_BT$ much lower than the bias energy $eV$ and the plasmon energy $\hbar\omega_0$, a condition which is fortunately realized experimentally.
\end{abstract}

\maketitle

\section{Introduction}

Electron transport through a nano system displays, due to the quantum nature of the underlying elementary processes, a current that exhibits quantum noise with zero-point fluctuations \cite{Blanter:00,Nazarov:02}. 
As a quantum object, the current is associated to a time-dependent operator $\hat{I}(t)$ in the Heisenberg representation. 
Hence, the noise spectral density $S(\omega)=\int dt e^{i\omega t} \langle \hat{I}(0)\hat{I}(t)\rangle$ acquires  a frequency-antisymmetric component $S(\omega)\neq S(-\omega)$ because of the noncommuting current operators at different times. 
This asymmetry can actually be accessed by coupling the system to a detector \cite{Lesovik:97,Gavish:00}. 
The result is that the positive and negative branches of $S(\omega)$ are related to the emission and absorption spectrum, respectively. 
Concerning the emission processes, if the source of noise is the system biased by a voltage $V$, intuitively one expects that the maximum energy available for the tunneling electron  is $eV$, and, thus, the energy of an emitted photon is limited to $eV$ as well, as shown by several experiments and theoretical investigations \cite{Lambe:76,Adams:81,Bloemer:87,Gornik:88,Gimzewski:88,Berndt:92,Berndt:01,Bharadwaj:11,
Rendell:78,Laks:79,Persson:92}. 
Such inelastic effects in tunneling junctions are interesting because they can help to reveal unusual phenomena like electron-electron correlation and electron-plasmon effects.

In regard to experimental measurements and realizations of current noise detection, one of the first proposals was a quantum tunneling detector consisting of a double quantum dot (DQD) coupled to the leads of a nearby mesoscopic conductor \cite{Aguado:00}, in which the inelastic current through the DQD measures the equilibrium and nonequilibrium fluctuations in the conductor \cite{Schoelkopf:02}.

Additionally, the light emission of electrons tunneling from a scanning tunneling microscope (STM) to a metallic surface has already been studied and used as a probe of the shot noise at optical frequencies for many years \cite{Berndt:90,Berndt:91,Ebbesen:03,Schneider:12,Berndt:15}.
%
%
%
%
%
%
%
\begin{figure}[tbp]
\includegraphics[width=0.99\columnwidth,clip=true]{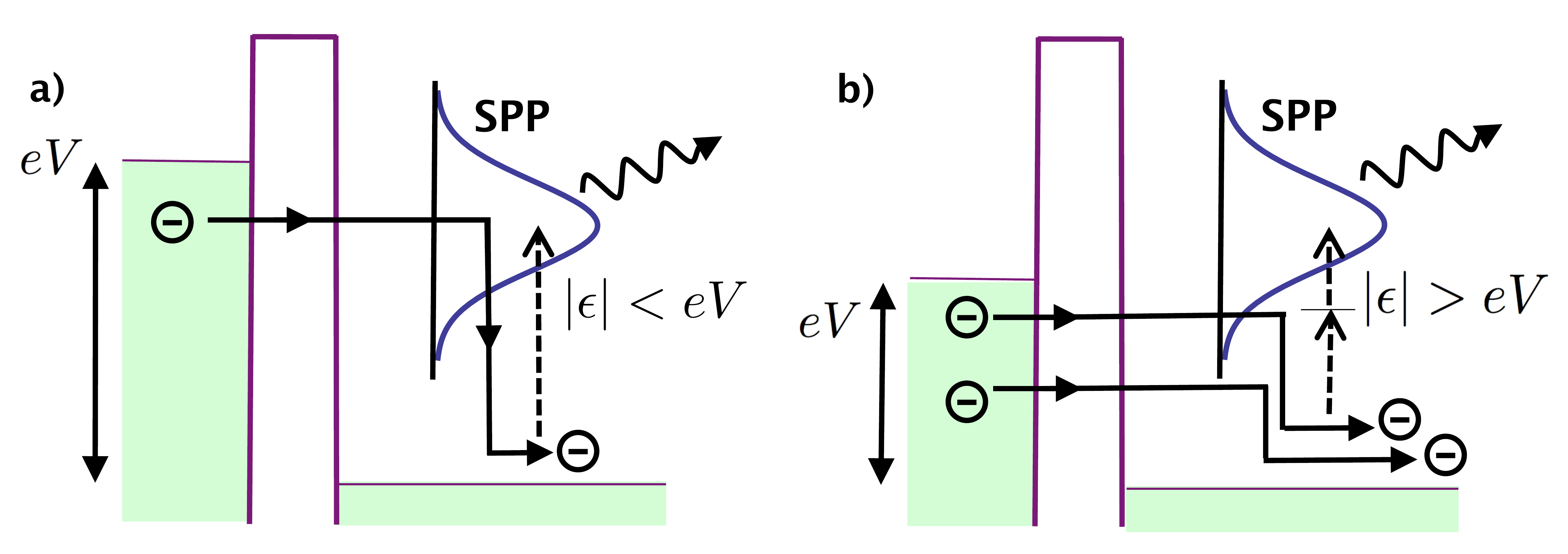}
\caption{Sketch of electron tunneling processes. (a) One electron tunnels through the barrier and excites the surface plasmon-polariton (SPP), which eventually emits a photon with energy $|\epsilon|<eV$. (b) The two coherent electrons tunnel through the barrier, creating an overbias SPP excitation and leading to the overbias light emission with energy $|\epsilon|>eV$.}
%
\label{fig:tunneling}
\end{figure}

Using a single electron scattering picture and at zero temperature of the system, the Pauli principle blocks  
inelastic tunneling transitions with energy exchange larger than the energy difference between the two Fermi seas, 
consisting of noninteracting electrons in the leads.
The emitted light spectrum is then limited in frequency by the bias voltage according to $\hbar\omega<eV$. i.e., the detector signals are in the sub-bias energy range 
$E<eV$; see Fig.~\ref{fig:tunneling}(a). 

However, some experiments \cite{Hoffmann:03,Schull:09,Schneider:10,Downes:02,Pechou:98,Gadzuk:71,Schneider:13} reveal the unexpected feature of light emitted at energy exceeding the bias voltage $\hbar\omega>eV$. 
Such an overbias spectrum appears as reminiscent of the surface plasmon-polariton (SPP) modes which can be also observed via other methods. 
Using essentially energy considerations, 
 such a process can be attributed to two simultaneously tunneling electrons providing enough energy to explain the overbias emission \cite{Fei:14,Kaasbjerg:15}.
%
%
Similar findings have also been reported in photon emission from Josephson junctions \cite{Hofheinz:11,Hassler:12,Blencowe:12,Armour:13,Kubala:14,Leppaekangas:15,Dambach:15} and molecular films \cite{Uemura:07,Fujiki:11,Dong:04,Dong:10} with fluorescent emission of photons with energies above the threshold energy. 
In order to understand these diverse systems, a detailed understanding of the electron tunneling processes involved is necessary \cite{Galperin:12}. 

In a previous Letter [\onlinecite{Fei:14}], we developed a theoretical framework for the description of the plasmon-mediated light emission 
by a tunnel contact based on dynamical Coulomb blockade.
%
In qualitative terms, in an elemental tunneling event, an electron gains energy $eV$ at bias voltage $V$ but must pay a charging cost of $E_c \sim e^2/C$ with $C$ the junction capacitance. Hence, after tunneling, a nonequilibrium situation occurs since the charge on the junction and the charge imposed by the voltage source are different. Now, when an impedance is connected in series to the tunnel junction, it allows us to discharge and dissipate energy, thus, reducing Coulomb charging effects. In other words, the electromagnetic environment of the junction crucially affects the charge tunneling events. The effect on the tunneling is captured by the probability $P(E)$ of emitting an energy amount $E$ to the electromagnetic environment. The so-called $P(E)$ function is related to the spectral density of voltage fluctuations, which in turn is determined by the impedance of the environment.\cite{Ingold:92,Devoret:90}

Going beyond the simple tunneling events, this framework captures the coherent two-electron tunneling processes, in which each electron contributes an energy $E_i \lesssim eV$ $(i=1,2)$ but 
the overall process creates an excitation in the broadened  SPP spectrum with an energy exceeding the bias voltage $E_1+E_2 > eV$,  as shown in Fig.~\ref{fig:tunneling}(b). 
 Afterwards, the relaxation of the SPP's energy finally leads to the overbias light emission.
By modeling the SPP as a broadened, damped resonator, at zero temperature we have quantitatively reproduced the experimentally observed bias-voltage-dependent emission spectrum [\onlinecite{Schull:09}].

Here, we extend our model described in Ref.~[\onlinecite{Fei:14}] to include a finite temperature in the general expression for the tunneling rate. 
First, we confirm that the non-Gaussian voltage fluctuations in the tunnel junction explain the light emission with energy above the bias voltage, $\hbar \omega > eV$, in the limit of low temperature.  
Second, we provide a quantitative estimation for the typical temperature above which overbias emission is masked by thermal effects.

Indeed, finite temperature affects either the rate associated to the Gaussian voltage fluctuations 
or the rate associated to the non-Gaussian voltage fluctuations. 
For the Gaussian rate, we find that increasing the temperature gradually smears out the sharp boundary at emission energy $E=eV$
which occurs in the limit of vanishing temperature.
For the non-Gaussian rate, finite temperature smooths the characteristic cusp of the overbias emission 
which is obtained at zero temperature.
Such effects are prominent even in the relatively low temperature regime, namely  $k_BT \sim 10^{-2} \hbar \Omega$ 
with $ \Omega \sim \omega_0$, the average position of the SPP spectrum, or $\Omega \sim \eta$, its broadening. 
These results point out that the overbias emission spectrum is sensitive to finite temperature effects.
However, remarkably, the non-Gaussian rate can still represent the leading term in the overbias range $E>eV$ for sufficiently low temperatures.
Hence, by analyzing the temperature dependence, the bias voltage dependence and their interplay for the individual rates, i.e. the 
Gaussian  and the non-Gaussian one, we discuss how to distinguish finite temperature effects from the expected 
``zero-temperature'' overbias emission. 

The structure of the paper is as follows. 
We describe the model and the theoretical methods based on the Keldysh action in Sec. II as well as the expression for the detection rate. 
In Sec. III, we calculate the total rate formed by two separate contributions, i.e., the Gaussian part and the non-Gaussian part, 
and analyze the rate behavior as a function of temperature and voltage bias. 
We discuss our conclusions in Sec. IV.
Details of the rate derivation are given in the Appendix.  

%
%
%
\section{Model}

\begin{figure}[tbp]
\includegraphics[width=0.99\columnwidth,clip=true]{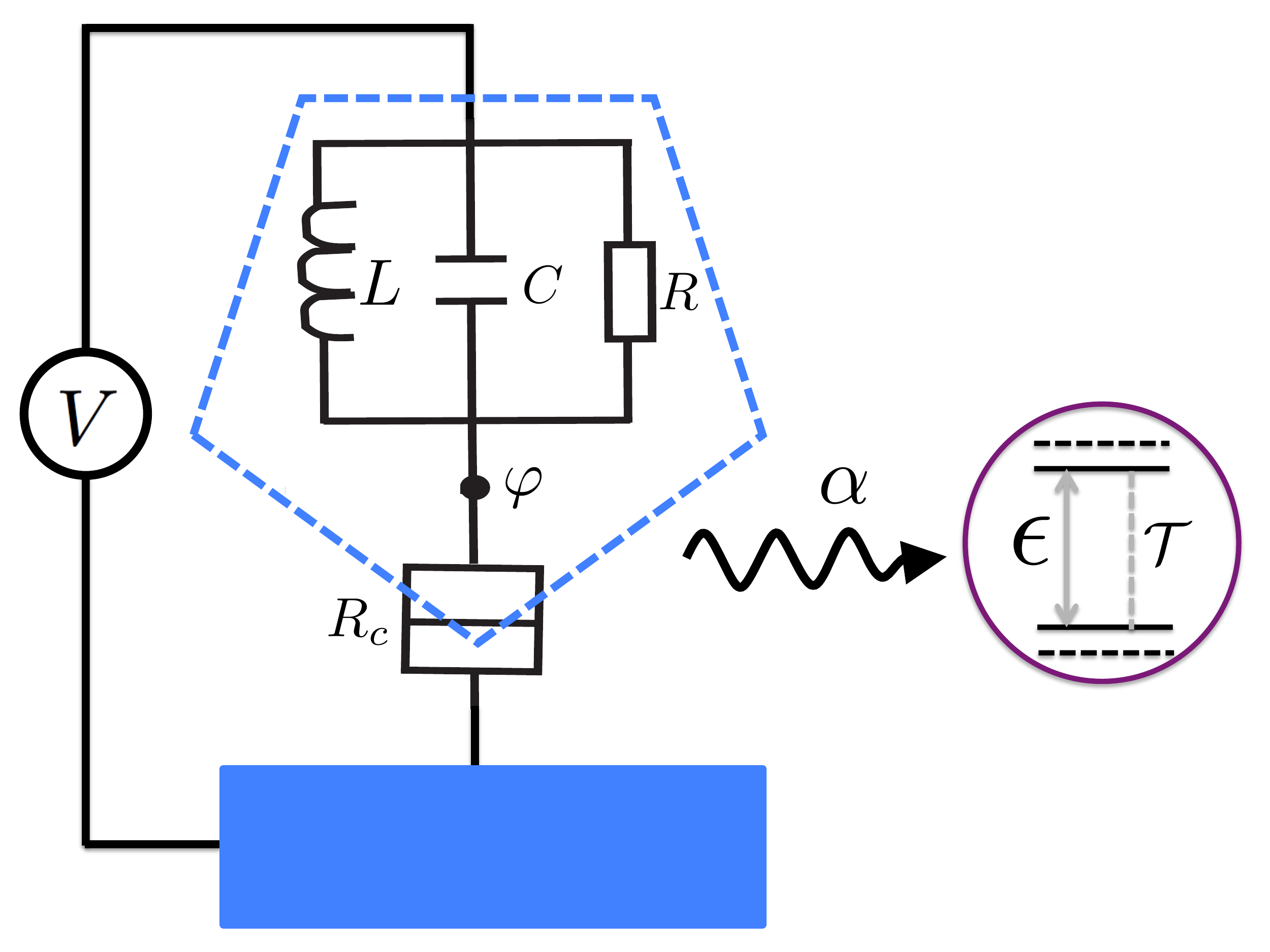}
\caption{Sketch of a STM contact with bias voltage $V$. The electrons interact via the SPP mode that is mimicked by the LRC resonant circuit. Photons emitted from this tunnel junction are absorbed by the detector, i.e. a two-level system with energy separation $\epsilon$, leading to the absorption process characterized by the transition probability $\mathcal{T}$. The coupling $\alpha$ between the detector and the tunnel system is weak in concordance with the experiment \cite{Schull:09}.}
\label{fig:overview}
\end{figure}

We model the tunneling from the STM tip to the surface in an electromagnetic environment, 
according to standard DCB theory \cite{Ingold:92,Devoret:90,Lesovik:97}, as the circuit diagram depicted in Fig.~\ref{fig:overview}.

The tunneling is described by a tunnel conductor that has a dimensionless conductance $g_{c}=R_{Q}/R_{c}$ with $R_Q=h/2e^2$ and $R_c$ being the quantum and tunneling resistances, respectively. 
The junction is coupled to a damped LC circuit, modeled by an impedance $z_\omega=iz_{0}\omega \omega_{0}/(\omega_{0}^2-\omega^2+i\omega \eta)$, where $\omega_0=1/\sqrt{LC}$ is the resonance frequency of the SPP mode, $\eta=1/RC$ models the damping, and $z_0=\sqrt{L/C}/R_{Q}$ is the scaled characteristic impedance.  
The interaction between the tunnel junction and the SPP occurs in this model via the dynamical voltage fluctuations on the node between the tunnel junction and the LRC circuit. 
These voltage fluctuations can be expressed by the phase variable $\varphi(t)=\frac{e}{\hbar}\int ^t_{-\infty} dt V(t')$.
%
%
%
%
%
%
%
%

For the photon detection, we choose a simple two-level system with energy difference $\epsilon$ and transition probability $\mathcal{T}$ to absorb or emit photons.  
Formally the system can be described by a Hamiltonian $H_{detec}=(\epsilon+\alpha V)\sigma_z/2+\mathcal{T}\sigma_x$ with the unperturbed eigenstates  $|\pm\rangle$ with energies $\pm\epsilon/2$, respectively. 
The coupling $\alpha$ between the STM junction and the photon detector is set to be weak, since the photon detectors in a typical experiment are far away from the junction. We can calculate the transition rate from the transition probability $P_{-\to+}(t)=|\langle -(t)|+\rangle|^2$ to lowest order in the coupling $\mathcal{T}$. 
Using Fermi's golden rule and setting $\hbar=1$, the transition rate at energy $\epsilon$  in the detector 
due to the fluctuations  of $\varphi(t)$ \cite{Tobiska:06, Ingold:92,Devoret:90} reads 
\begin{equation}
	\label{eq:rate}
	\Gamma(\epsilon)=\lvert \mathcal{T} \rvert ^2 \int dt 
	\langle e^{i\alpha \varphi(t)} e^{-i\alpha \varphi(0)} \rangle e^{i\epsilon t} \, .
\end{equation} 
This rate formula corresponds to emission or absorption for $\epsilon>0$ and $\epsilon<0$, respectively.
In this work we study only the absorption rate, and therefore we have to consider only negative energies $\epsilon<0$.
Our central theoretical task is the calculation of $\Gamma(\epsilon<0)$ to the lowest order in the detector coupling constant, i.e., $\alpha$.

We employ the path integral method to evaluate $\langle e^{i\alpha \varphi(t)}e^{-i\alpha \varphi(0)}\rangle$.
Using the Keldysh actions of the conductor, $\mathbb{S}_{c}$, and the circuit, $\mathbb{S}_{e}$,
the correlator can be represented as
\begin{eqnarray}
	\langle e^{i\alpha \varphi(t)}e^{-i\alpha \varphi(0)}\rangle=& \int
	\mathcal {D}[\Phi]\exp \{
	-i\mathbb{S}_{e}[\Phi]-i\mathbb{S}_{c}[\Phi]  \nonumber \\
	&+i\alpha[-\varphi^{+}(0)+\varphi^{-}(t)]\},
\end{eqnarray}
where the two-component phase $\Phi= (\phi,\chi)^T $ with $\phi=(\varphi^++\varphi^-)/2$ and $\chi=\varphi^+-\varphi^-$,
 and the real fields $\varphi^\pm(t)$ are defined on the forward and backward Keldysh contours, respectively. 
Later in the Keldysh action, the real fields can be written as $\varphi_{\omega}^{\pm}=\phi_{\omega}\pm \frac{1}{2}\chi_{\omega}$ 
in frequency space.

The action of the damped LC oscillator acting as the environment on the tunnel conductor, is quadratic in the fields 
\cite{Kindermann:03,Kamenev:01} and given by 
\begin{equation}
\mathbb{S}_{e} \!=\! \int \!\!  \frac{d\omega}{2\pi} \,  \Phi^{T}_{-\omega}A_\omega\Phi_{\omega}\,
\, , \,
A_\omega\!=\!-\frac{i}{2} \! \left( \!\! \begin{array}{cc} 0 &
-\frac{\omega}{z_{-\omega}} \\
\frac{\omega}{z_{\omega}} & W(\omega) \Re \{
\frac{1}{z_{\omega}} \} \end{array} \!\! \right)  , \nonumber 
\end{equation}
with $W(\omega)=\omega \coth (\omega/2T)$. Here $T$ denotes the temperature and we have set $k_B=1$.
The action $\mathbb{S}_{c}$ can be expressed in terms of Keldysh Green's functions $\check{G}_{L,R}$ for the free electrons on the left ($L$) and right ($R$) sides of the tunneling barrier \cite{Belzig:01}:  
\begin{equation}
\label{eq:Sc} 
\mathbb{S}_{c}=\frac{i}{8} g_{c}\int dtdt'\mathrm{Tr}\{ \check{
G}_{L}(t,t'), \check{G}_{R}(t'-t) \} \,
\end{equation}
in the tunneling limit $g_c \ll 1$ \cite{notes} .
%
%
%
With  the help of the equilibrium Keldysh Green's function
\begin{equation}
\label{eq:G_omega}
	\check{G}(\omega)=\left(
	\begin{array}{cc}
	1-2f(\omega) & 2f(\omega) \\
	2[1-f(\omega)] & 2f(\omega)-1
	\end{array}\right) ,  
\end{equation} 
containing the Fermi function $f(\omega)=[\exp(\omega/T)+1]^{-1}$, we can write $\check G_R(\omega)=\check G(\omega-eV)$ and hence $\check G(t)=\int d\omega \exp(-i\omega t)\check G(\omega)/2\pi$.
Again using the Fourier transformation, we write $\check{G}_{L}(t,t^{\prime }) =\check U^\dagger(t) \check{G}(t-t^{\prime })\check U(t')$ with the counting fields introduced as 
\begin{equation}
\label{eq:U_t}
\check U(t)=\left(
\begin{array}{cc}
e^{-i\varphi ^{+}(t)} & 0 \\
0 & e^{-i\varphi ^{-}(t)}
\end{array}
\right). 
\end{equation}
Due to the nonquadratic contribution to the action of the conductor $\mathbb{S}_{c}$ in Eq.~(\ref{eq:Sc}),  the correlator cannot be calculated exactly and we need an approximation scheme. Here, we use the cumulant expansion for the action $\mathbb{S}_{c}$ by which we obtain the result
\begin{equation}
\Gamma(\epsilon)=\Gamma_{\mathrm{G}}(\epsilon)+\Gamma_{\mathrm{nG}}(\epsilon)+\mathcal{O}(\lambda^2) \, .
\end{equation}
The first Gaussian term scales as $\Gamma_{\mathrm{G}}(\epsilon) \sim \Gamma_{0}= (2 \pi)^2 \alpha^2 {|\mathcal{T}|}^2 g_{c} z_{0}^2 /\omega_{0}$ whereas the second non-Gaussian terms scales as $\Gamma_{\mathrm{nG}} \sim g_{c} z_{0}^2 \Gamma_{0}$ pointing out
that the validity of our expansion is based on the smallness of the expansion parameter $\lambda=g_{c}z_{0}^2$.

%
%
%
\section{Results}

%
%
%
%
\subsection{Gaussian contribution}

A first approximation is obtained by considering only the quadratic part of the conductor action, in which the whole path integral becomes Gaussian 
and, in the limit of vanishing voltage $V=0$, corresponds to the well-known results from P(E) theory. 
The quadratic part of the conductor action reads
\begin{equation*}
	\mathbb{S}_{c}^{\mathrm{G}} 
	\!=\! \int \!\!  \frac{d\omega}{2\pi} \,\,   
	 \Phi_{-\omega}^{T} B_{\omega} \Phi_{\omega} \,\,,\,\, 
	B_\omega= -\frac{i}{2} \left( \begin{array}{cc}
	0 & -\omega g_{c} \\
	\omega g_{c} & S_{c}(\omega)
	\end{array} \right),
\end{equation*} 
with the symmetrized quantum noise of a tunnel contact $S_{c}(\omega)=\frac{1}{2} g_{c} (W(\omega+eV)+W(\omega-eV))\equiv g_{c}\widehat{W}(\omega)$. At $T=0$ temperature, the symmetrized quantum noise vanishes for $|\omega|>eV$ thus we can already conclude that, even if just the Gaussian part of the conductor action is included, Eq.~(\ref{eq:rate}) can only describe photon emission with energies limited by the bias voltage.

Combining all the quadratic parts from both the LRC circuit and the conductor in a single matrix,
\begin{equation*}
 D_{\omega} \equiv \frac{1}{2\pi}(A_{\omega}+B_{\omega})
=-\frac{i}{4\pi} \left( \begin{array}{cc}
0 & -\frac{\omega}{\tilde{z}_{-\omega}} \\
\frac{\omega}{\tilde{z}_{\omega}} & S(\omega)
\end{array} \right),
\end{equation*}
 with $S(\omega)=S_{c}(\omega)+W(\omega) \Re\{ 1/z_{\omega}\}$. Then, the correlation function $\langle e^{i\alpha \varphi(t)}e^{-i\alpha \varphi(0)} \rangle\equiv e^{\alpha^2J(t)} $ can be calculated. As a result, one finds
\begin{eqnarray}
	J(t) =  
	\int d\omega \frac{\lvert \tilde{z}_{\omega} \rvert ^{2}}{\omega^{2}}
	 S_t(\omega)(e^{-i\omega t}-1),
\end{eqnarray}
where 
\begin{eqnarray*}
S_t(\omega)&=&2\pi (S(\omega)+\omega \Re \{ 1/\tilde{z}_{\omega}\}) \\
&=& 2\pi g_{c} [\widehat W(\omega)+\omega] +2\pi [W(\omega)+\omega]\Re\{ 1/z_{\omega}\} 
\end{eqnarray*}
is the total nonsymmetrized noise spectral density. 
The impedance $\tilde{z}_{\omega}=z_{\omega}/(1+z_{\omega}g_{c})$ is the parallel connection of the tunnel junction and the environmental impedance playing the role of the  ``effective environment" to the detector. 
This means that the factor $g_c$ leads to an increased damping of the resonator, which can be absorbed in a renormalized $\eta\to\eta+1/R_cC$ and will be ignored henceforth.
From Eq.~(\ref{eq:rate}), in the lowest order in $\alpha^2$, we obtain the rate in scaled units, 
\begin{equation}
	\label{eq:rategauss}
	\Gamma_{\mathrm{G}}(\epsilon) \, =
	2\pi \alpha^2 {\left| \mathcal{T} \right|}^2 \frac{ {\left| \tilde{z}_{\epsilon} \right|}^2  }{\epsilon^2} S_{t}(\epsilon) \, ,
\end{equation}
which is consistent with the known emission rate at finite temperature.

\begin{figure}[t]
\centering
\includegraphics[width=0.8\columnwidth,clip=true]{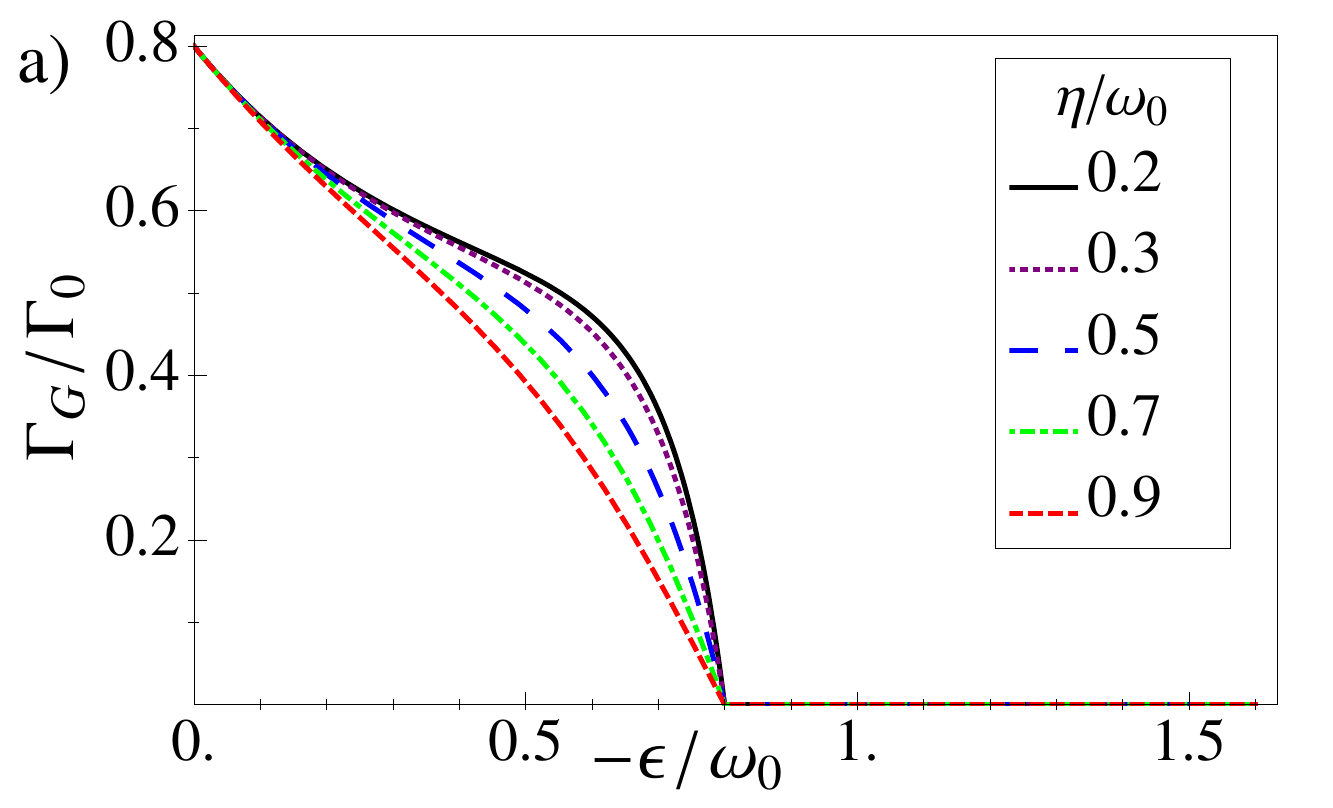} \\
\includegraphics[width=0.8\columnwidth,clip=true]{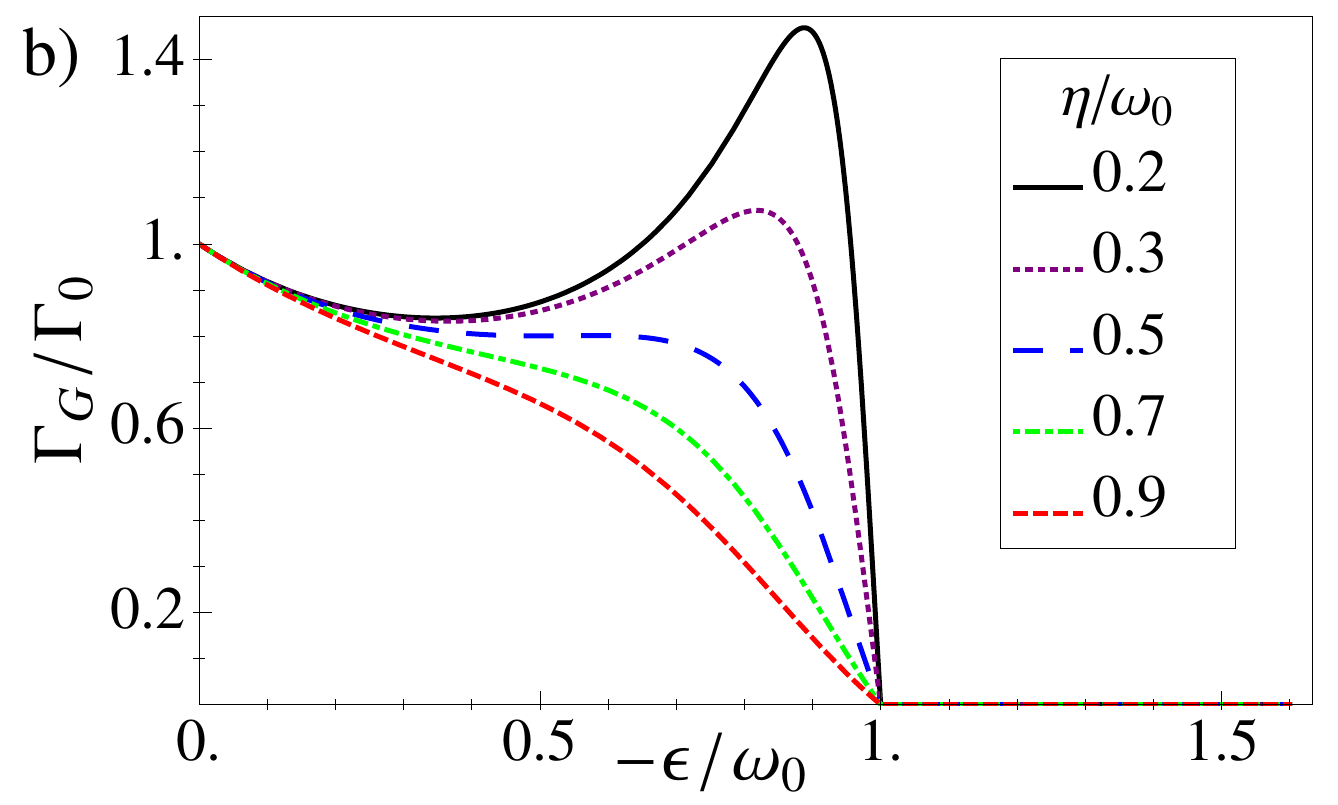} \\
\includegraphics[width=0.8\columnwidth,clip=true]{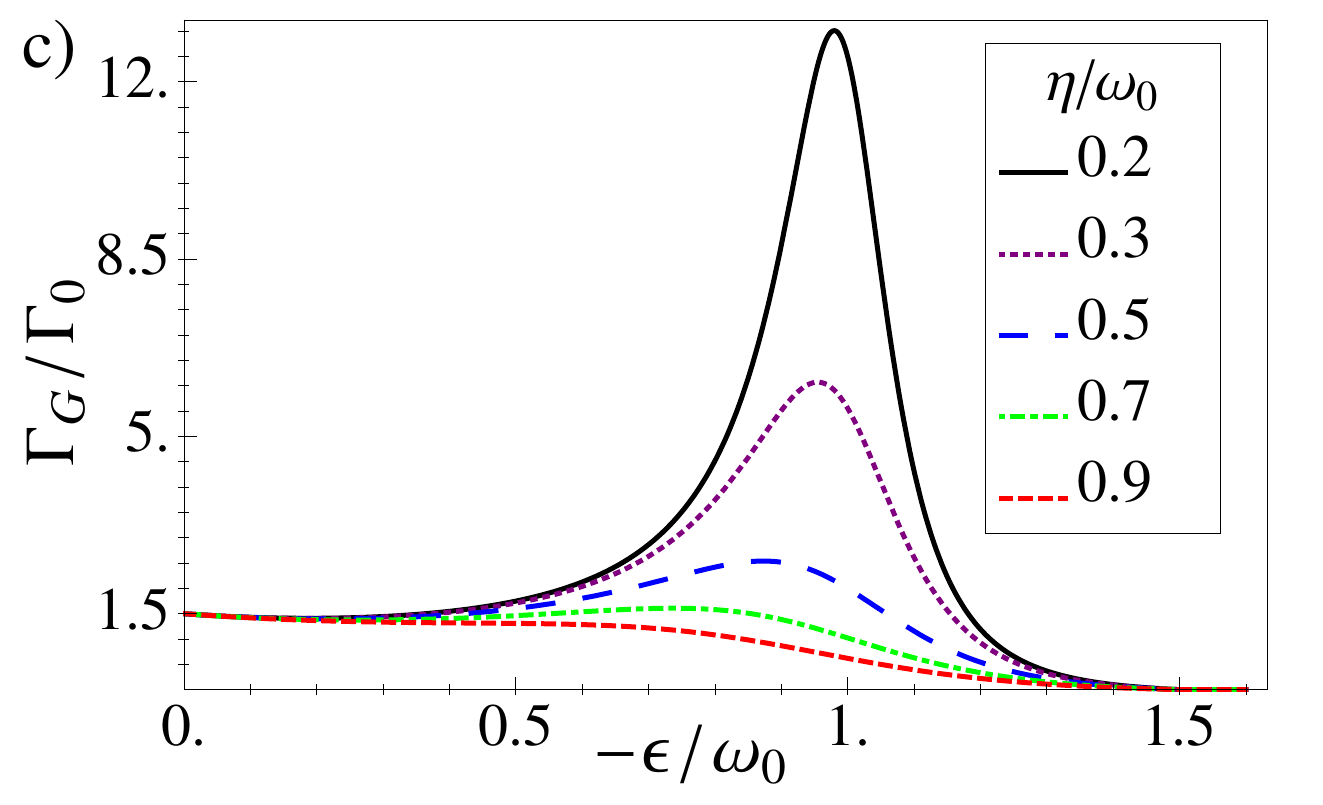}
\caption{The Gaussian contribution to the emission spectrum for different broadenings, at zero temperature for three different values of the bias voltage:  
(a) voltage below the resonance $eV=0.8\omega_0$,
(b) voltage at the resonance $eV=\omega_0$, 
(c) voltage above the resonance $eV=1.5\omega_0$.  In all cases, the threshold occurs at $|\epsilon|=eV$. 
The SPP resonance becomes visible once the threshold is larger than $\omega_{0}$. 
The smaller the broadening $\eta$ is, the sharper the SPP peak becomes. }
\label{fig:T0Gauss}
\end{figure}

In the limiting case $T \rightarrow 0$, $W(\omega) \rightarrow |\omega|$ 
and the result (\ref{eq:rategauss}) reduces to the one obtained in Ref.~[\onlinecite{Fei:14}], namely 
\begin{equation}
	\label{eq:rategauss_T=0}
	\Gamma_{\mathrm{G}}(\epsilon) \, = \, 
	 	 \Gamma_0  \, 
		 R_{\eta} \left( \epsilon \right) \, 
		 \theta \left(eV + \epsilon \right)  \, 
		  \left( \frac{eV + \epsilon  }{\omega_0} \right) 
		  \quad \left( T=0 \right) \, ,
\end{equation}
in which we set the dimensionless resonance shape function $R_{\eta}(\epsilon) = 1/[ {(\epsilon^2/\omega_0^2-1)}^2+ \epsilon^2 \eta^2/\omega_0^2]$. 
In this limit the maximum energy $eV$ for a photon emission due to inelastic transitions  is $eV$ as a consequence of the 
sharp Fermi surfaces on both sides of the tunnel junction, and the emission spectrum 
has indeed a cutoff at  $|\epsilon|=eV$. 
In Fig.~\ref{fig:T0Gauss}  we give an example of the Gaussian emission spectrum at zero temperature for three different values of the bias voltage at different damping parameters $\eta$.  
%
%
%
%
%
%
%
%
%
%
At a voltage below the resonance $eV<\omega_0$  in Fig.~\ref{fig:T0Gauss}(a), the broadening has only a small influence on the emission spectrum and no peak occurs in the spectrum.
The SPP resonance is visible only when the bias voltage becomes comparable or larger than the resonance $\omega_{0}$, 
as shown in Figs.~\ref{fig:T0Gauss}(b) and \ref{fig:T0Gauss}(c).
For instance, in Fig.~\ref{fig:T0Gauss}(b), close to the threshold $eV$ the emission is enhanced, but the threshold remains clearly visible.  
In the limit of large bias voltage $eV>\omega_0$ [Fig.~\ref{fig:T0Gauss}(c)], the full resonance is reflected in the emission spectrum 
and its shape is essentially determined by the resonance function appearing as a prefactor to the noise
in Eq.~(\ref{eq:rategauss_T=0}).  
Hence, the maximum is $\sim 1/\eta^2$ and can be strongly increased in high-quality resonators 
or well-defined plasmonic modes.

%
%
%
%
\begin{figure}[t]
\centering
\includegraphics[width=0.9\columnwidth,clip=true]{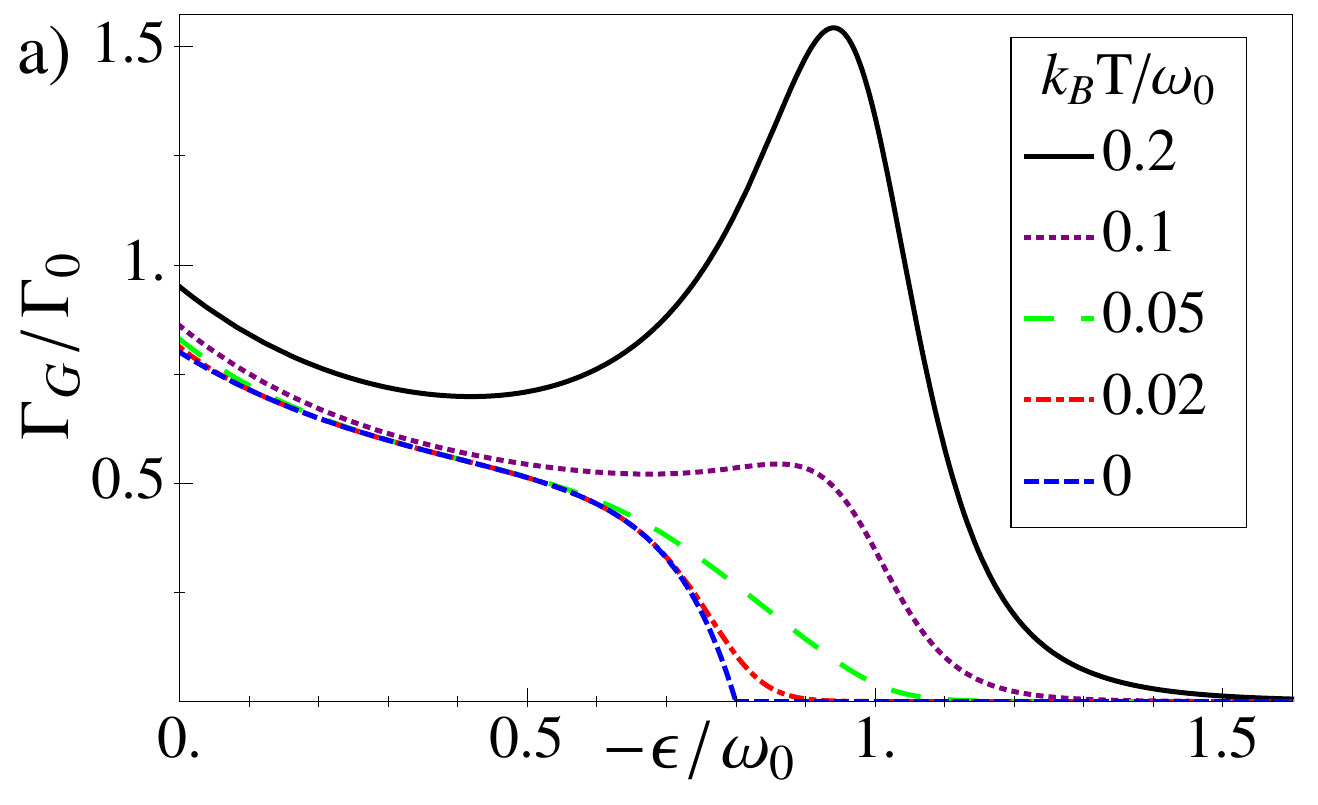} \\
\includegraphics[width=0.9\columnwidth,clip=true]{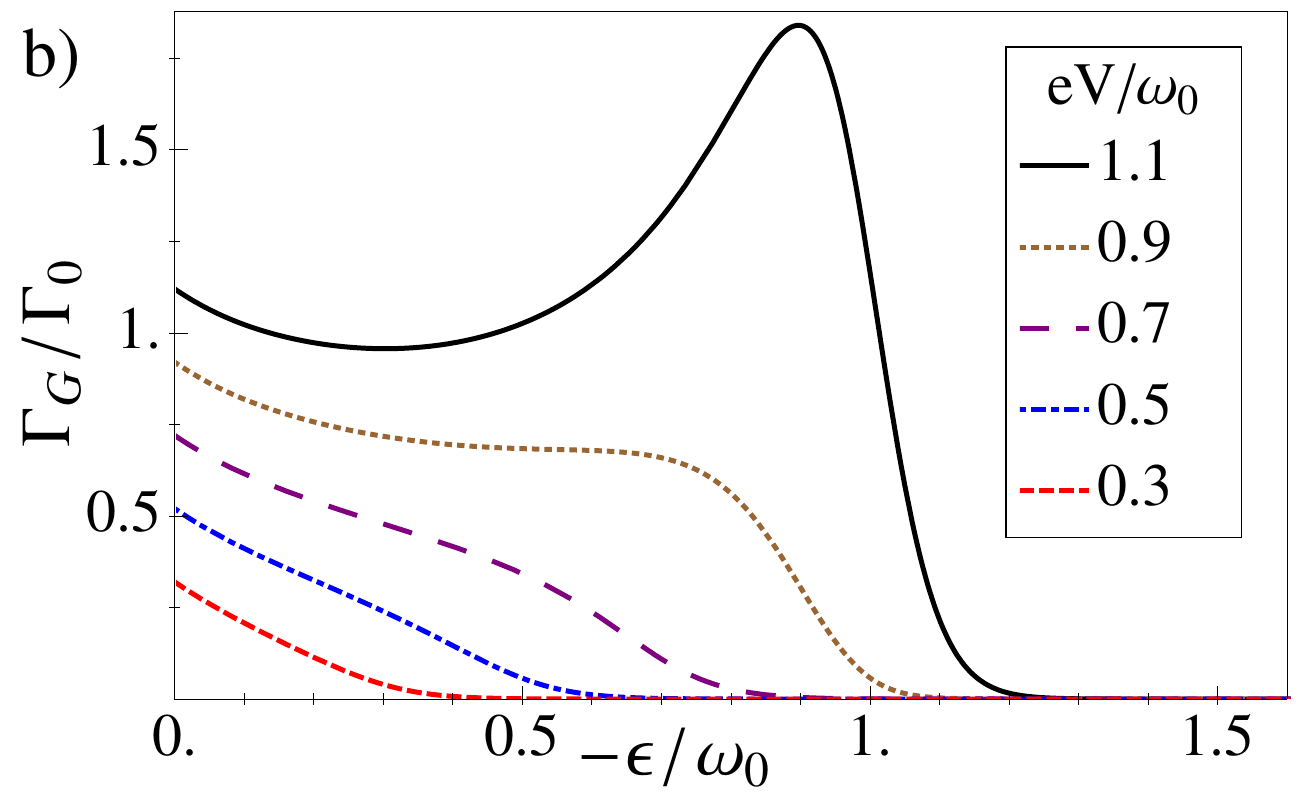} 
\caption{(a) The Gaussian contribution to the emission spectrum for different temperatures with the bias voltage $eV/\omega_{0}=0.8$. As the temperature is increased, the cutoff at the bias $eV$ is washed out. In addition, more electrons are involved in the tunneling processes, leading to an increased rate. (b) The Gaussian contribution to the emission spectrum for different bias voltages. The SPP peak is more pronounced when the bias voltage exceeds the resonance energy $\omega_{0}$. In all cases, the sharp threshold for $-\epsilon>eV$ that exists at zero temperature is smoothened at finite temperature, which is already achieved at the surprisingly small but finite temperature $T= \omega_{0}/30$. The broadening parameter in both figures is chosen as $\eta=0.3 \omega_{0}$, whereas the product of the tunneling conductance and 
the characteristic impedance of the resonator is set to  $g_c z_0 =1$.}
\label{fig:Gauss}
\end{figure}

At finite temperature we can cast the Gaussian rate as 
\begin{equation}
	\label{eq:rategauss_T_finite}
	\Gamma_{\mathrm{G}}(\epsilon) = 
	 	 \Gamma_0  R_{\eta} \left(  \epsilon \right)  
		\left[ \! \frac{\widehat{W}(\epsilon)+\epsilon}{\omega_0} \!+\!  
		\left(\!\frac{1}{g_c z_0}\!\right)  \frac{ W(\epsilon)+\epsilon}{\omega_0} \! \right] 
		  \, , 
\end{equation}
and the clear cutoff at $T=0$ due to the Fermi distribution is smoothed out.  

Figure~\ref{fig:Gauss}(a) shows the emission rate for different temperatures at a voltage below the resonance $eV=0.8\omega_0$ and for $g_c z_0 =1$.
Different values of the ratio $g_c z_0$ do not change the result significantly provided that $eV \gg T$ because  
the noise of the intrinsic thermal contribution of the plasmon - corresponding to the second term in Eq.~(\ref{eq:rategauss_T_finite}) - 
scales as $~\exp[-eV/T]$ around the cutoff $|\epsilon|=eV$ and it is hence exponentially small.
Since a finite temperature softens the sharp cutoff at $|\epsilon|=eV$ that exists at zero temperature, the SPP resonance can come into play even 
at an energy lower than the bias voltage, thus contributing an overbias emission as well. 
The resonance strongly enhances the thermally excited overbias emission. 
It is remarkable that the step at $eV$ is already almost invisible at a small temperature of just a few \% of $\omega_0$. 
This can be traced back to the thermally excited quasiparticles in the lead with the higher chemical potential 
 - corresponding to the first term in Eq.~(\ref{eq:rategauss_T_finite}) - 
 so that the thermal tail at the resonance is 
 $\sim \exp[-(\omega_0-eV)/T]$ with $\omega_0 \sim eV$ and therefore exponentially larger than the intrinsic thermal contribution of the plasmon.

Figure~\ref{fig:Gauss}(b) shows the emission rate for different  bias voltages at low temperature $T=\omega_0/30$. 
In all cases, from bias voltages below the resonance $eV<\omega_0$ to bias voltages above the resonance $eV>\omega_0$, 
we have the disappearance of the zero temperature cutoff at $|\epsilon|=eV$.
As long as the voltage becomes larger than $eV>\omega_0$, the SPP resonance becomes visible in the emission spectrum in a similar way 
to the case of vanishing temperature $T=0$. 
In other words, at finite and small temperatures $T \ll \omega_0$, 
we have substantial corrections to the zero-temperature result for the Gaussian rate  
around the cutoff at $|\epsilon|=eV$.

\subsection{Non-Gaussian contribution}

Although single-electron tunneling events produce signatures of the overbias SPP peak at finite temperature,
we will now turn to the nonquadratic part of the action $\mathbb{S}_{c}$ describing the electron-electron correlation 
that gives contributions to the overbias emission. 
As pointed out in Ref.~[\onlinecite{Fei:14}], comparing the absolute orders of magnitude, 
the non-Gaussian phase fluctuations are smaller than the dominating Gaussian fluctuations 
due to the small environmental impedance $g_c  z_{\omega}^2  \ll 1$.
However, the non-Gaussian rate represents the only one contribution to the total rate in the overbias region $|\epsilon| \gg eV$ 
at $T=0$. 
We aim to understand in which range of parameters, for sufficiently low temperature and well inside the overbias region $|\epsilon| > eV$,  
the non-Gaussian rate can continue to dominate over the thermal Gaussian rate. 
Before discussing the results for the non-Gaussian rates, we report the main steps for calculating such a rate. Further details are given in the Appendix. 

First, from Eqs.~(\ref{eq:Sc})-(\ref{eq:U_t}),   we expand the action of the coherent conductor to the fourth order of $\varphi$
while the higher-order terms can be neglected due to the factor  
$g_{c}z_{\omega}^2 \ll 1$, yielding $\mathbb{S}=\mathbb{S}_{e}+\mathbb{S}_{c}^G+\mathbb{S}_{c}^{(3)}+\mathbb{S}_{c}^{(4)}+\mathcal{O}(\Phi^5)$. 
Second, using $\exp [-i \mathbb{S}_{c}^{(3)} -i \mathbb{S}_{c}^{(4)}] \approx 1 -i \mathbb{S}_{c}^{(3)} -i\mathbb{S}_{c}^{(4)}$, 
in accordance with the approximation above, we can write the path integral as 
\begin{equation}
        \label{eq:expansion}
	\langle e^{i\alpha \varphi(t)}e^{-i\alpha \varphi(0)}\rangle \simeq \, 
	 e^{\alpha^2 J(t)} \!\! 
	-  i \langle \langle \, \mathbb{S}_{c}^{(3)}  \rangle \rangle  \, 
	-  i \langle \langle \, \mathbb{S}_{c}^{(4)}  \rangle \rangle,  
\end{equation}
in which we used the Gaussian average $\langle \langle \cdots \rangle \rangle \equiv \int \mathcal{D}[\Phi](\cdots)e^{\int d\omega \{ -i \Phi^{T}_{-\omega}D_{\omega}\Phi_{\omega}+i\alpha b^{T}_{\omega}(t) \Phi_{\omega}\} }$ and  $b_{\omega}(t)=(e^{-i \omega t}-1,-(e^{-i \omega t}+1)/2)^T$. 
After expanding for small $\alpha$, the first term  in Eq.~(\ref{eq:expansion}) yields the Gaussian rate discussed in the previous section.
Concerning $\mathbb{S}_{c}^{(3)}$, it is an odd term which gives a nonvanishing result 
only to the order $\alpha^3$ and we neglect it for $\alpha\ll1$. 
Thus, we focus on the fourth term which  is given in frequency space by  
%
%
%
\begin{displaymath}
\begin{aligned}
&\mathbb{S}^{(4)}_{c}= \frac{1}{12} \frac{1}{(2\pi)^{4}} \frac{i}{8} g_{c}\int d\omega d\omega' d\omega''  \\
\{\! & \Big(\!2[F(\omega)+F(-\omega)]-3 [F_{1}(\!-\omega\!-\!\omega ')\!+\!F_{2}(\!-\omega\!-\!\omega ')]\!\Big) \\
& \ [\varphi^{+}(\omega')\varphi^{+}(\omega) \varphi^{+}(\omega'')\varphi^{+}(-\omega-\omega'-\omega'') \\
& \ \ +\varphi^{-}(\omega')\varphi^{-}(\omega) \varphi^{-}(\omega'')\varphi^{-}(-\omega-\omega'-\omega'')] \\
-& 4F(-\omega)\varphi^{+}(\omega)\varphi^{-}(\omega')\varphi^{-}(\omega'')\varphi^{-}(-\omega\!-\!\omega'\!-\!\omega'') \\
-& 4F(\omega)\varphi^{-}(\omega)\varphi^{+}(\omega')\varphi^{+}(\omega'')\varphi^{+}(-\omega\!-\!\omega'\!-\!\omega'') \\
+&  6F_{1}(-\omega\!-\!\omega')\varphi^{+}(\omega)\varphi^{+}(\omega')\varphi^{-}(\omega'')\varphi^{-}(-\omega\!-\!\omega'\!-\!\omega'') \\
+&6F_{2}(-\omega\!-\!\omega')\varphi^{-}(\omega)\varphi^{-}(\omega')\varphi^{+}(\omega'')\varphi^{+}(-\omega\!-\!\omega'\!-\!\omega'') \} \, ,
\end{aligned}
\end{displaymath}
%
%
%
%
with
\begin{displaymath}
\begin{aligned}
&F_{1}(\omega)=(-\omega-eV)+W(-\omega-eV), \\
&F_{2}(\omega)=(\omega+eV)+W(\omega+eV), \\
&F(\omega)=F_{1}(\omega)+F_{2}(-\omega) \, .
\end{aligned}
\end{displaymath}
For the field $\varphi^{\pm}_{\omega}$, we list the results:
\begin{displaymath}
\begin{aligned}
&\langle \langle \Phi_{\omega} \rangle \rangle = \langle \langle \begin{array}{l}
\phi_{\omega} \\
\chi_{\omega} \end{array} \rangle \rangle=\frac{1}{2}D^{-1}_{\omega}b_{-\omega}(1+\mathcal{O}[\alpha^{2}z^{2}]) \\
&=2\pi i \alpha \begin{pmatrix}
\bigg [ S(\omega)\frac{\lvert \tilde{z}_{\omega}\rvert^{2}}{\omega^{2}}-\frac{1}{2}\frac{\tilde{z}_{\omega}}{\omega} \bigg ] e^{i\omega t}- \bigg [S(\omega)\frac{\lvert \tilde{z}_{\omega}\rvert^{2}}{\omega^{2}}+\frac{1}{2}\frac{\tilde{z}_{\omega}}{\omega} \bigg ] \cr
-\frac{\tilde{z}_{-\omega}}{\omega}e^{i\omega t}+\frac{\tilde{z}_{-\omega}}{\omega}
\end{pmatrix}
\\
&\textrm{and}
\ \langle \langle \Phi_{\omega} \Phi_{-\omega}^{T} \rangle \rangle =
\left(\begin{array}{cc}
\langle\langle \phi_{\omega}\phi_{-\omega} \rangle\rangle &
\langle\langle \phi_{\omega}\chi_{-\omega} \rangle\rangle \\
\langle\langle \chi_{\omega}\phi_{-\omega} \rangle\rangle & 
\langle\langle \chi_{\omega}\chi_{-\omega} \rangle\rangle
\end{array} \right)
\\
&= -\frac{i}{2}D_{\omega}^{-1} =2\pi \left( \begin{array}{cc}
S(\omega) \frac{\lvert \tilde{z}_{\omega} \rvert^{2}}{\omega^{2}} & \frac{\tilde{z}_{\omega}}{\omega} \\
-\frac{\tilde{z}_{-\omega}}{\omega} & 0 
\end{array} \right).
\end{aligned}
\end{displaymath}
%
%
%
%
%
%
In the weak coupling limit, $\alpha \ll 1$, corresponding to weak detection that is the experimentally relevant regime, the
main order pairings of averages appearing in $\mathbb{S}^{(4)}_{c}$ are  of the type
$\langle \langle \varphi_{\omega} \rangle \rangle \langle \langle \varphi_{-\omega} \rangle \rangle \langle \langle \varphi_{\omega'}\varphi_{-\omega'}\rangle \rangle$
and they are proportional to $ \sim \alpha^2$.
Such terms can be calculated using Wick's theorem to find all possible pairings of single and double averages.
Finally, we consider only the lowest order terms in $\sim g_{c}^2$ in order to obtain the following expression for the non-Gaussian contribution:
\begin{widetext}
\begin{eqnarray}
\label{eq:nG-rate}
&\Gamma_{\mathrm{nG}} = &
\frac{\pi g_{c}^2 \alpha^2 {|\mathcal{T}|}^2 }{2}\frac{|\tilde{z}_{\epsilon}|^2}{\epsilon^2}\! \! \int_{0}^{\infty} \! d\omega 
\Bigg\{\frac{|\tilde{z}_{\omega}|^2}{\omega^2}
\left(\widehat{W}(\omega)-W(\omega)\right)
\Big[-2\widehat{W}(\epsilon)+\Big(\widehat{W}(\omega+\epsilon)+\widehat{W}(\omega-\epsilon)\Big) \Big]  
\nonumber \\
&& +2
\left(\widehat{W}(\epsilon)-W(\epsilon)\right) 
\frac{Re \{ \tilde{z}_{\epsilon}\}}{\epsilon}\frac{Re \{ \tilde{z}_{\omega}\}}{\omega} \Big[\widehat{W}(\omega+\epsilon)-\widehat{W}(\omega-\epsilon) \Big]  \nonumber \\
 &&+2
 \left(\widehat{W}(\epsilon)-W(\epsilon)\right) 
 \frac{ Im\{ \tilde{z}_{\epsilon}\}}{\epsilon} \frac{Im \{ \tilde{z}_{\omega}\}}{\omega}\Big[ 2W(eV)-2\widehat{W}(\omega)-2\widehat{W}(\epsilon)+\widehat{W}(\omega+\epsilon)+\widehat{W}(\omega-\epsilon) \Big]\!\Bigg \} \, .
\end{eqnarray}
\end{widetext}
More details on the deriviation of this expression can be found in the Appendix.

\begin{figure}[h!]
\centering
\includegraphics[width=0.8\columnwidth,clip=true]{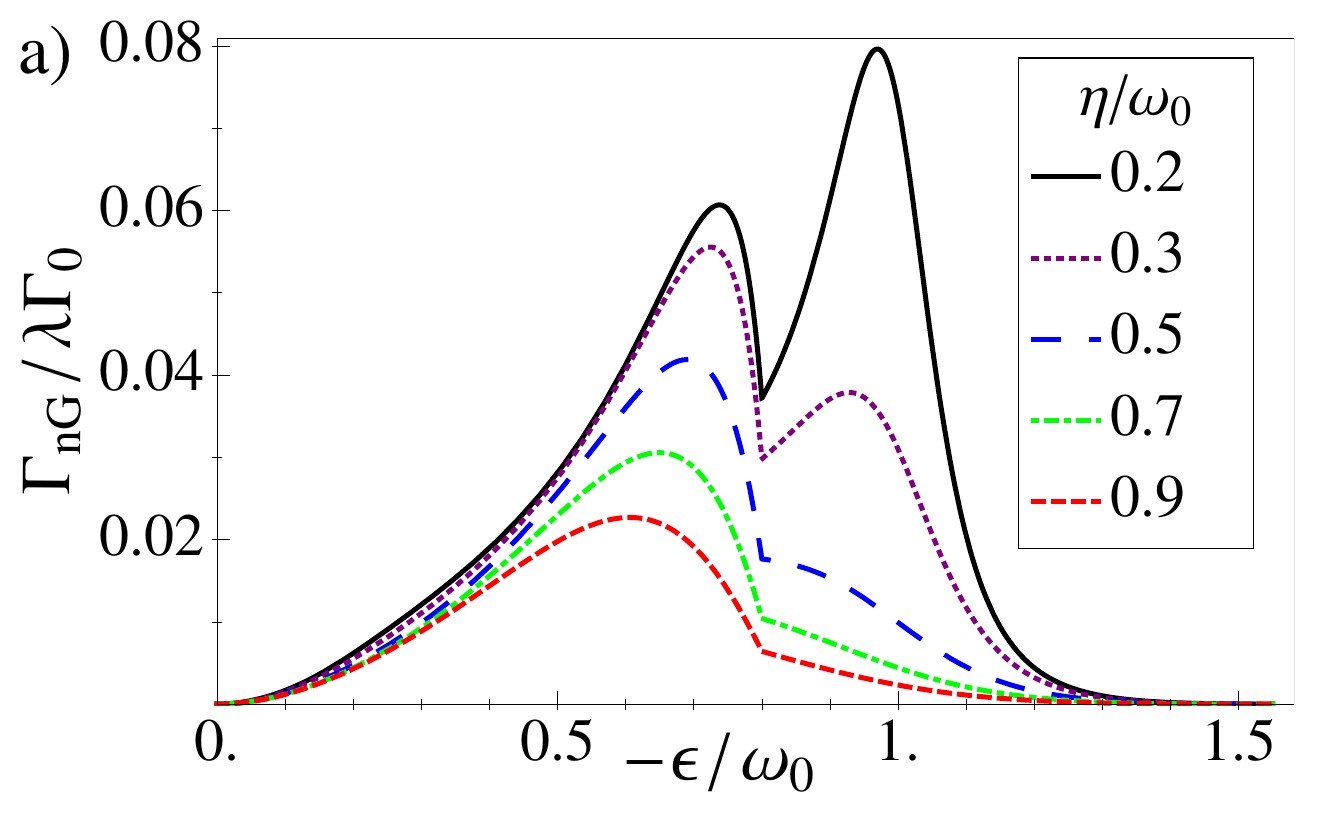} \\
\includegraphics[width=0.8\columnwidth,clip=true]{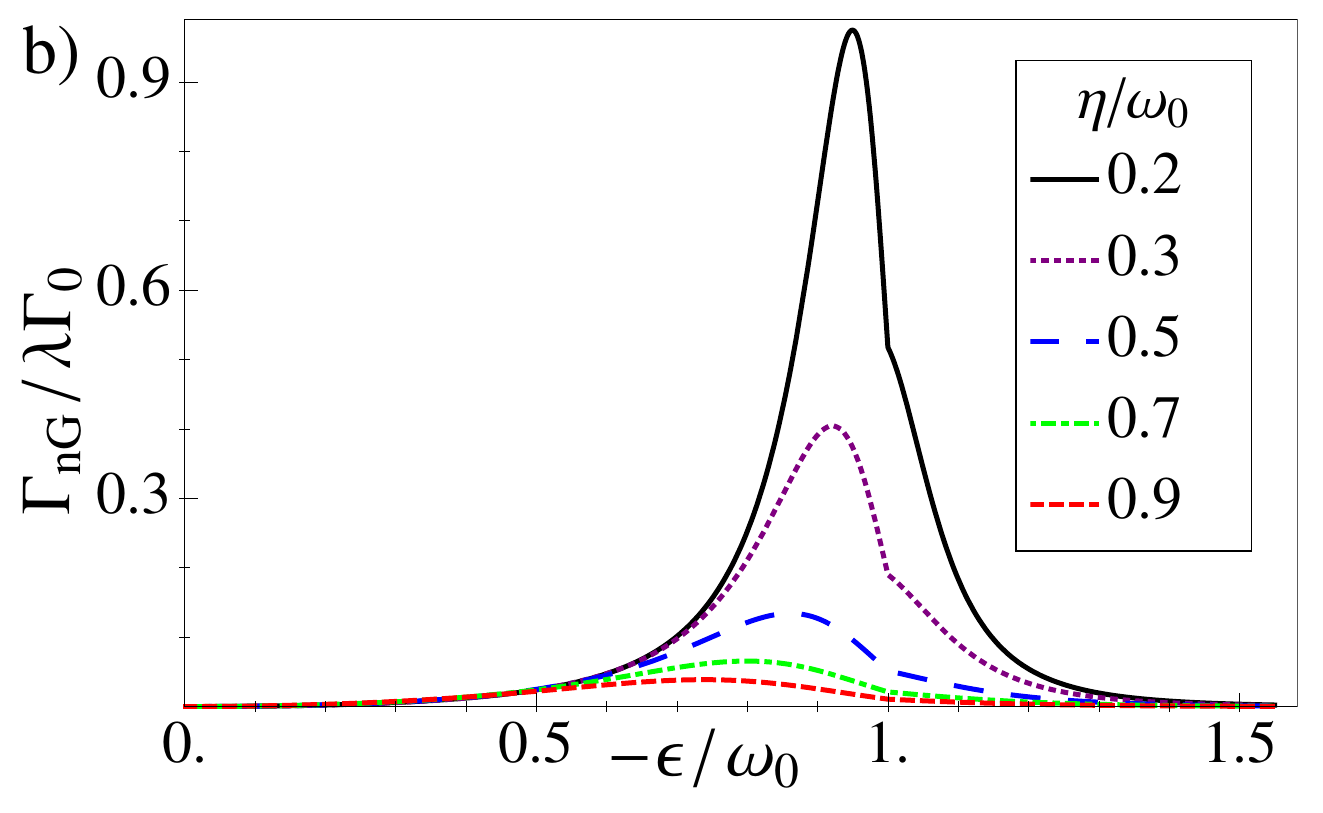} \\
\includegraphics[width=0.8\columnwidth,clip=true]{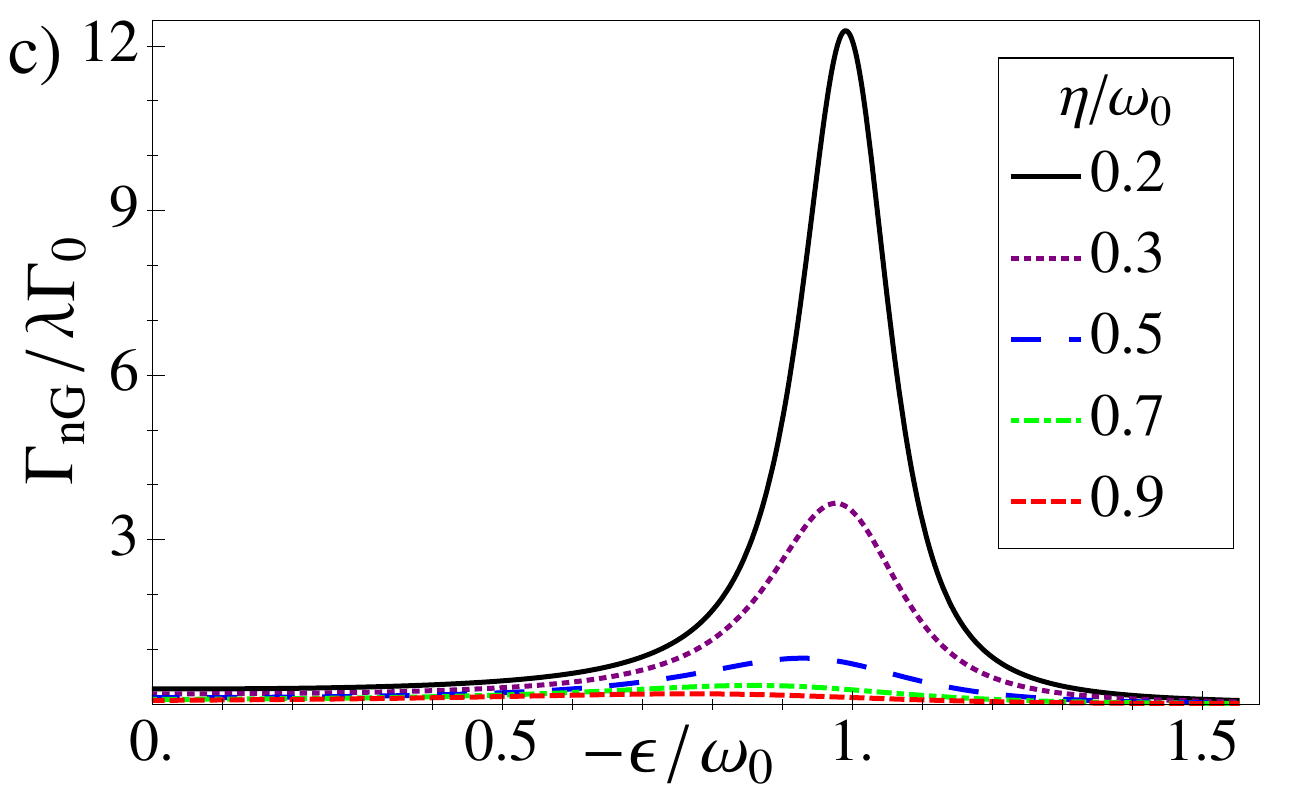}
\caption{The non-Gaussian contribution at zero temperature to the emission spectrum for different broadenings. There is a kink for 
$|\epsilon|=eV$ whereas the resonance peak appears always at  $|\epsilon|=\omega_0$. 
Parameters are the same as in Fig.~\ref{fig:T0Gauss}.}
\label{fig:T0nGauss}
\end{figure}

Similarly as for the Gaussian fluctuations, we recover our former results [\onlinecite{Fei:14}] in the limit $T\rightarrow 0$.
Examples of the non-Gaussian rate at zero temperature are given in Fig.~\ref{fig:T0nGauss}  scaled with $\lambda \Gamma_0$ 
and with $\lambda=g_c z_0^2$, our expansion parameter. 
%
%
%
%
%
%
%
%
The non-Gaussian rate yields a contribution in the underbias as well as in the overbias regime.
Moreover, the non-Gaussian rate here calculated to lowest order in $\alpha$ and $g_c$ 
has also a high-energy cutoff at $|\epsilon| =2 eV$ above which  $\Gamma_{\mathrm{nG}}=0$.
The latter result is in agreement with the picture of two correlated electrons involved in a single photon emission 
whose energy is now limited by $\hbar\omega<2 eV$.  
Such a cutoff is less pronounced  than the sharp cutoff of the Gaussian rate 
at $|\epsilon|=eV$ although it is evident in the experimental data (see next section and Fig.~\ref{fig:fitting}).

As for the Gaussian case in Fig.~\ref{fig:T0Gauss}, we plot in Fig.~ \ref{fig:T0nGauss} the three different cases corresponding to 
bias voltages below or above the resonance $eV<\omega_0$ or $eV>\omega_0$, and the resonant case $eV=\omega_0$.

In the first case $eV<\omega_{0}$, Fig.~ \ref{fig:T0nGauss}(a), the curve for the non-Gaussian rate shows a characteristic cusp 
at the threshold $|\epsilon|=eV$.
Such a curve has also peaks in both the underbias region $|\epsilon|<eV$ as well as in the overbias region  $|\epsilon|> eV$ 
in correspondence with the resonance of the SPP mode  at  $|\epsilon|=\omega_0$. The overbias emission at $T=0$ corresponds to the first line of Eq.~\ref{eq:nG-rate}.
However, in the underbias region $|\epsilon|<eV$, the non-Gaussian rate is dominated by the 
leading Gaussian contribution  so that the first peak hardly can be distinguished 
and one expects that the overbias emission rate is distinctly resolved around the resonance $\omega_0> eV$ only.

For bias voltages at the resonance $eV=\omega_0$ , Fig.~ \ref{fig:T0nGauss}(b), the two peaks associated with 
the non-Gaussian rate  merges into a single peak and the curve shows a kink at the threshold $|\epsilon|= eV$. 
In this case the non-Gaussian rate has still a  noticeable contribution in the overbias regime  $|\epsilon| > eV$ in terms of  
the tail of the resonance peak centered at the threshold.

Then, for the last case, $eV>\omega_0$, shown in Fig.~ \ref{fig:T0nGauss}(c), the non-Gaussian rate behaves in a way similar 
to the Gaussian rate in Fig.~ \ref{fig:T0nGauss}(c) with a single peak at the resonance $|\epsilon| = \omega$.
Such a peak is now located well inside the underbias region in which the non-Gaussian rate is dominated by the Gaussian rate.

\begin{figure}[t]
\centering
\includegraphics[width=0.8\columnwidth,clip=true]{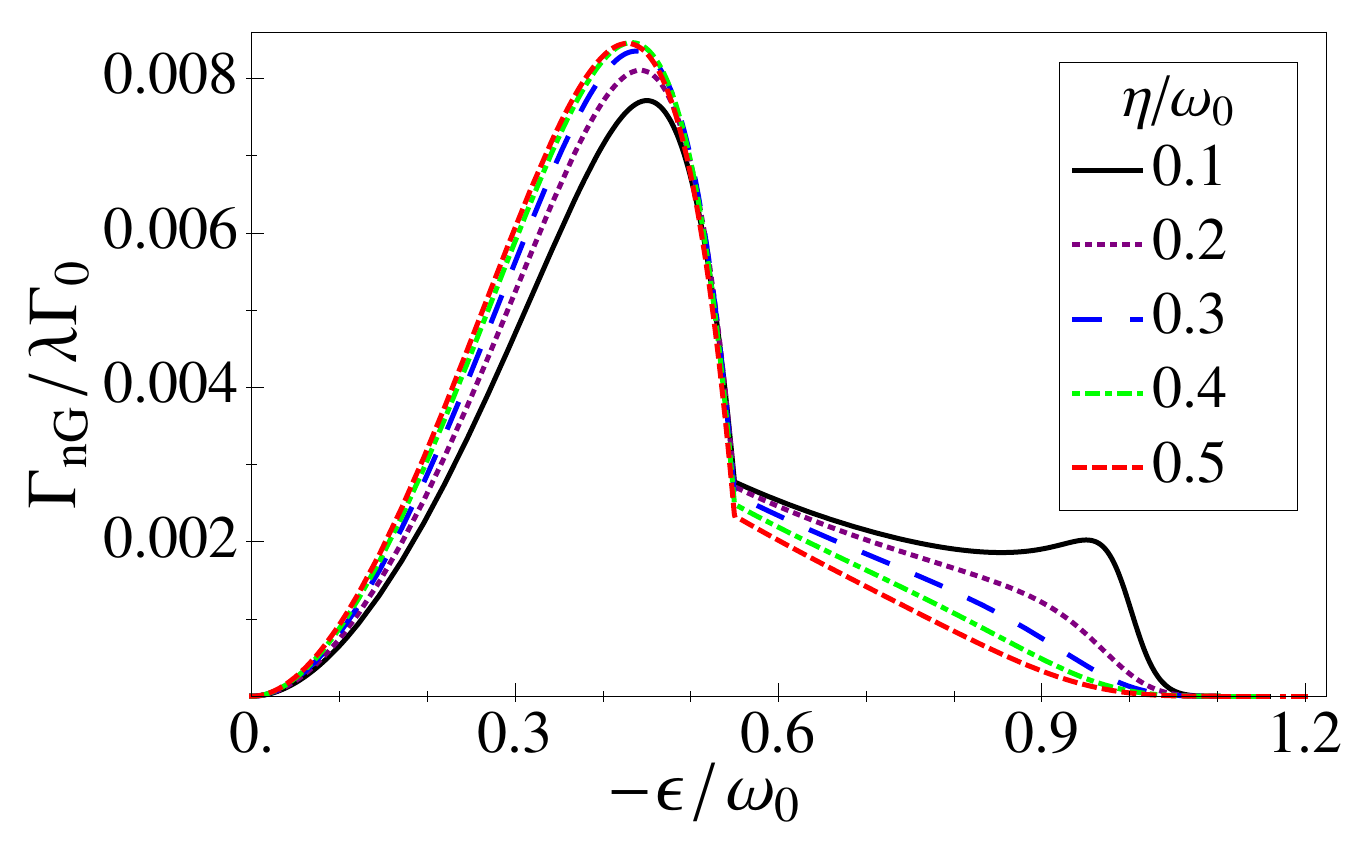} \\
\caption{The non-Gaussian contribution at zero temperature to the emission spectrum for different broadenings. 
The bias voltage is $eV=0.55\omega_0$ such that the two-electron energy cutoff is at $2eV=1.1\omega_0$.}
\label{fig:T0nGaussV055}
\end{figure}

Finally, we consider the case that when the SPP resonance $\omega_{0}$ is quite close to the two-electron energy cutoff $2eV$, which is shown in Fig.~\ref{fig:T0nGaussV055}.  
Here we can see, unlike Fig.~\ref{fig:T0nGauss}(a) where the SPP resonance $\omega_{0}$ is far away from the $2eV$ cutoff, that the overbias peak can still be present although strongly weakened.

Thus we can conclude that overbias photon emission due to the non-Gaussian voltage fluctuations in mesoscopic tunnel junctions 
is, \textit{a priori}, always a possible effect even far away from the resonance of the plasma-polariton modes, but the effect's magnitude 
can be smaller than the limit of a photon detector. 
On the contrary, the overbias photon emission becomes a substantial effect provided that the system has a resonant plasmonic mode 
at a frequency in the overbias range $eV \geq \omega_0 $ and below the cutoff for the two electrons emission $\omega_0 < 2eV$. 

%
%
%
%
%
%
%
%

\begin{figure}[th]
\centering
\includegraphics[width=0.8\columnwidth,clip=true]{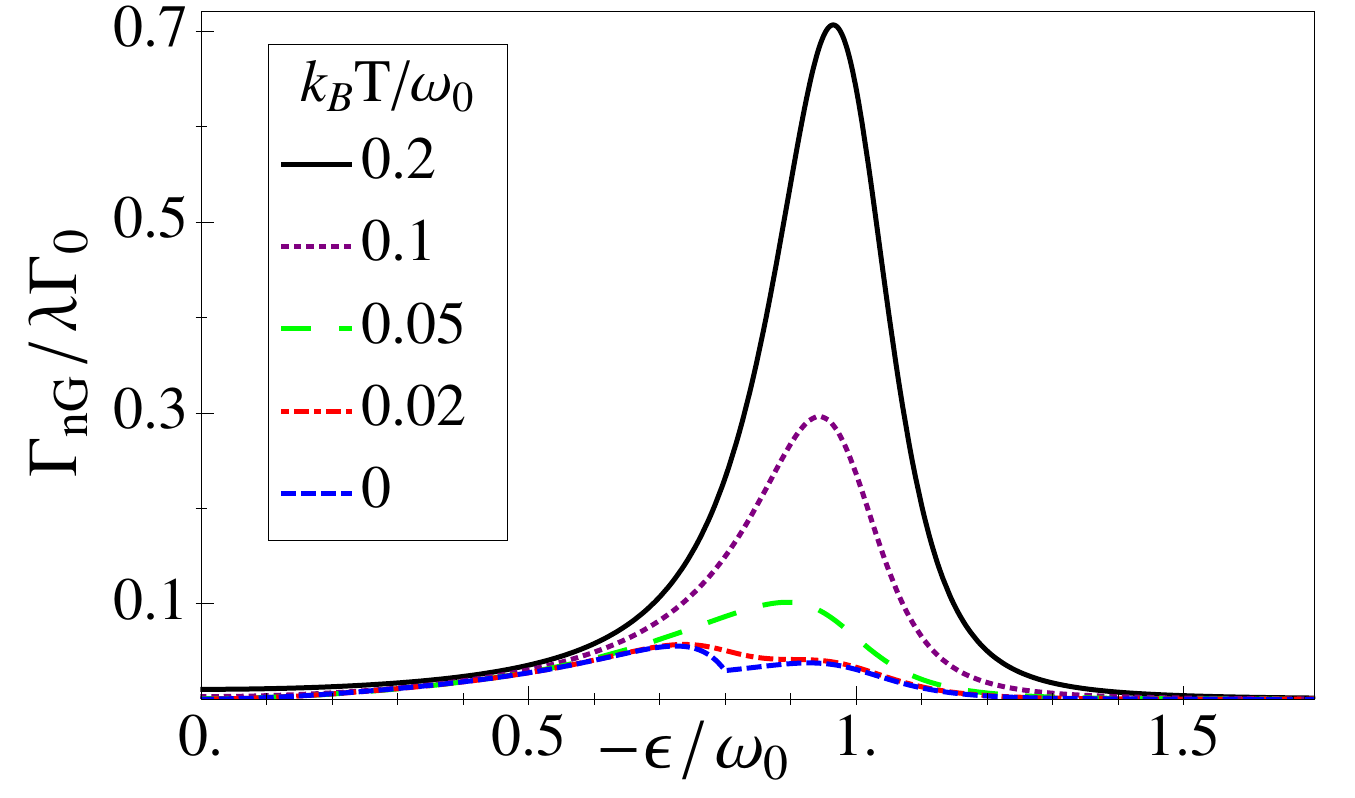}
\caption{The non-Gaussian contribution to the emission spectrum for different temperatures at the bias voltage $eV/\omega_{0}=0.8$ 
and $\eta=0.3 \omega_{0}$.
Due to the increased temperature, the kink at the bias disappears and the two peaks are merged into a single peak.}
\label{fig:nGausswithT}
\end{figure}

We discuss now the effects of a finite temperature for the non-Gaussian rate for the case $eV<\omega_0$. 
Some examples are shown in Fig.~\ref{fig:nGausswithT} with an intrinsic broadening of the SPP mode $\eta=0.3\omega_0$. 
%
%
%
%
%
%
%
%

In order to distinguish between the low and high temperature regimes, \textit{a priori} we can compare the broadening $\eta$ 
with the thermal smearing expected at finite temperature $\sim k_BT$. 
Then one expects that the non-Gaussian rate continues to exhibit sharp features in the low temperature range, defined by $k_B T < \eta$ 
and that it becomes a smooth, smeared function as the temperature approaches the broadening $k_B T \lesssim \eta$. 
In Fig.~\ref{fig:nGausswithT},  we can see that, increasing the temperature, the two distinct peaks merge into a single peak
and the kink at the bias voltage $|\epsilon|=eV$ is weakened concealing any overbias signatures. 
Remarkably, this merging occurs even at relatively low temperature $T\sim 10^{-2} \omega_0$ 
compared to the broadening of the mode $\eta \sim 10^{-1}  \omega_0$ pointing out that 
the overbias is highly sensitive to finite temperature.

On the other hand, increasing the temperature enhances the height of the peak in a similar way as the Gaussian rate, as discussed in the previous section.
In other words, above the threshold $|\varepsilon|>eV$ and at finite temperature, one can not discriminate the overbias emission due to the Gaussian fluctuations 
 - associated to single electron processes - from the  overbias emission due to the non-Gaussian fluctuations - associated to two-electron processes.
In order to resolve such processes, we have to consider the low temperature range.
%
%
%
%
%
%
%
\begin{figure}[t]
\centering
\includegraphics[width=0.8\columnwidth,clip=true]{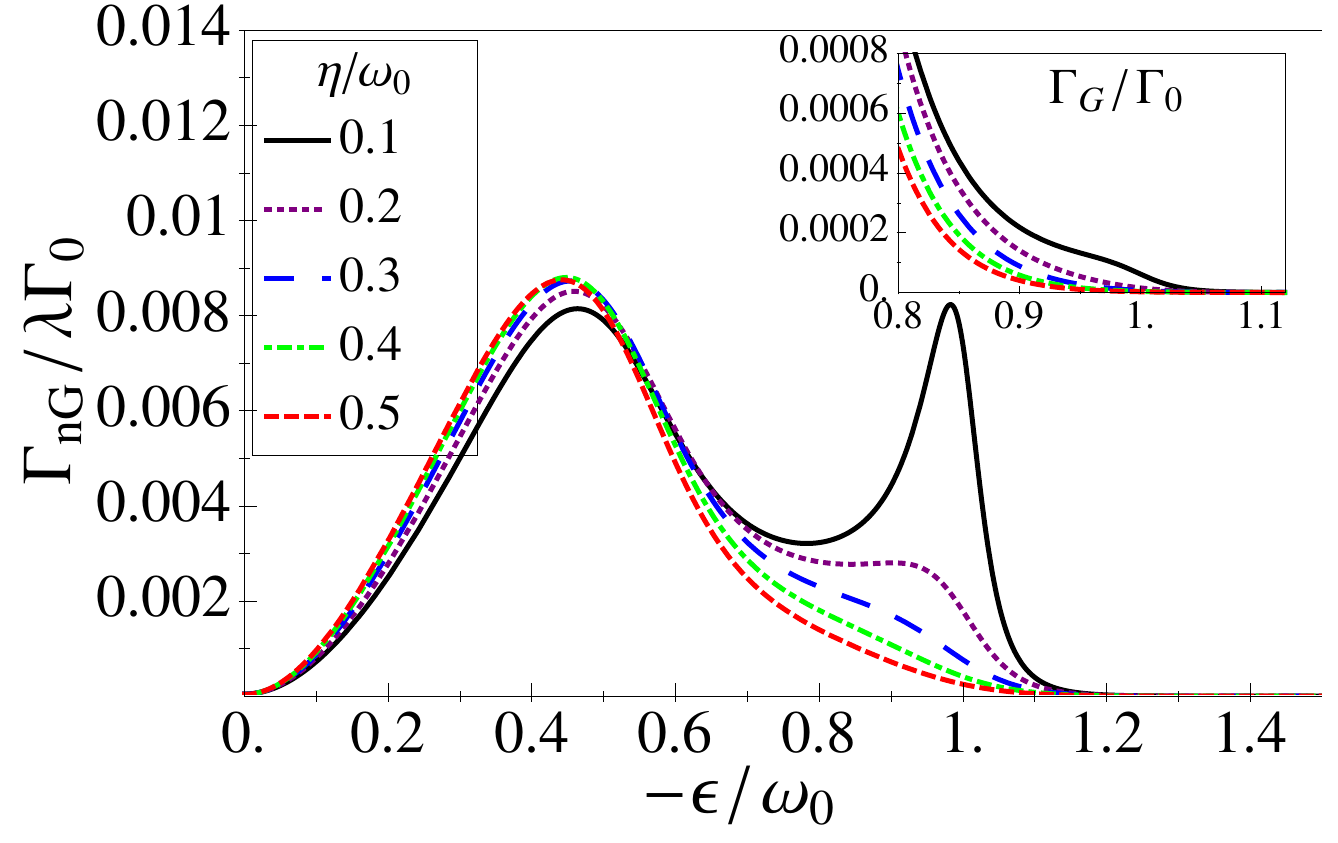}
\caption{The non-Gaussian contribution at finite temperature $T=\omega_0/30$ to the emission spectrum for different broadenings,  
at $2eV=1.1\omega_0$, viz. the SPP resonance dominates near the $2eV$ cutoff.
The inset shows the Gaussian contribution around the SPP resonance.
Thus, in this case, with proper $\lambda$, even at finite temperature, the overbias due to the two-electron emission (non-Gaussian part) can dominate the Gaussian one. }
\label{fig:nGaussV055}
\end{figure}

In Fig.~\ref{fig:nGaussV055}, we discuss the behavior of the non-Gaussian rate at low temperature, $T=\omega_0/30$, 
as varying the damping $\eta$ when the resonance is  close to the two-electron cutoff $2eV=1.1 \omega_0$.
By comparing with the Gaussian part under the same condition - the inset of Fig.~\ref{fig:nGaussV055} - 
we notice that at finite temperature, since the bias voltage $eV$, that is important for the single electron emission, is far away from the resonance, the Gaussian part around the resonance is small as it is due to the temperature smearing of the Fermi distribution. 
Meanwhile, the non-Gaussian part can represent the larger contribution in the case of a sharp resonance.
%

%
%
%
%

\subsection{Total rate and Comparison with the experiments}

\begin{figure}[hbtp!]
\centering
\includegraphics[width=0.8\columnwidth,clip=true]{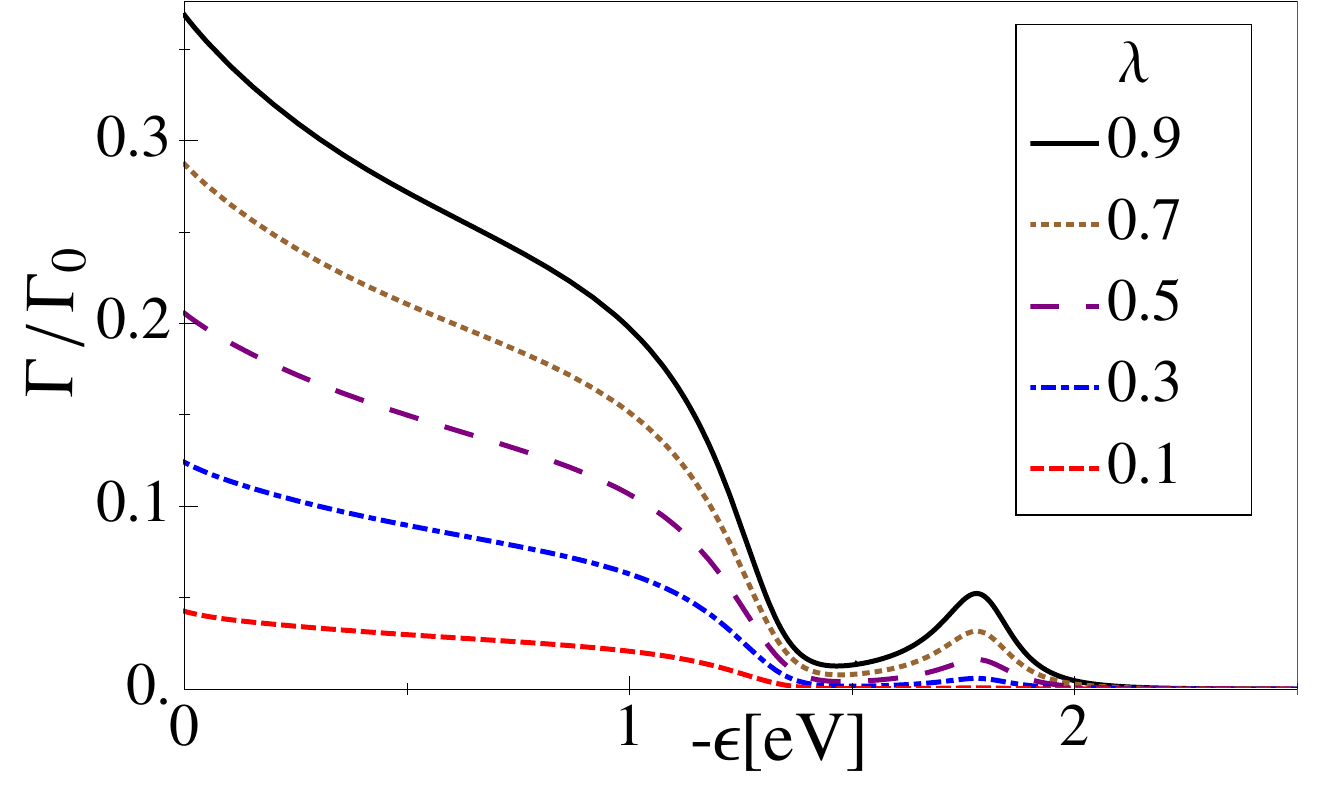}
\caption{The total rate for different dimensionless factors $\lambda=g_{c}z_{0}^2$ at the bias voltage  
$V=1.32$ V. The overbias peak increases with increasing $\lambda$, which determines the weight of the non-Gaussian part to the total rate. The temperature is chosen to be the room temperature $\beta \omega_{0}=\omega_{0}/k_{B}T=72$ and the SPP resonance energy is taken to be $\omega_{0}=1.8$ eV with the broadening $\eta=0.2$ eV. }
\label{fig:with gc}
\end{figure}

For the total tunneling rate, we have to take the Gaussian as well as the non-Gaussian rates into account.
In order to compare the theoretical results with the experimental data of  G. Schull and co-workers [\onlinecite{Schull:09}], 
in this section we plot the rate explicitly as a function of energy $(eV)$ for a SPP mode centered at 
$\omega_0= 1.8$ eV and broadening $\eta=0.2$ eV.
As is known from Eq. (5) and Eq. (6), these two rates are normalized by a dimensionless factor of $\lambda=g_{c}z_{0}^2$. 
Then as $\lambda$ increases, the non-Gaussian rate gradually gives the dominant contribution to the total emission rate in 
the overbias energy regime, leading to the overbias emission peak becoming more visible (see Fig.~\ref{fig:with gc}).
However, for small $\lambda$, within the validity of our expansion, the non-Gaussian features are weak and smeared out by the Gaussian properties due to the finite temperature.
%
%
%
%
%
%
%
%

\begin{figure}[t!]
\centering
\includegraphics[width=0.8\columnwidth,clip=true]{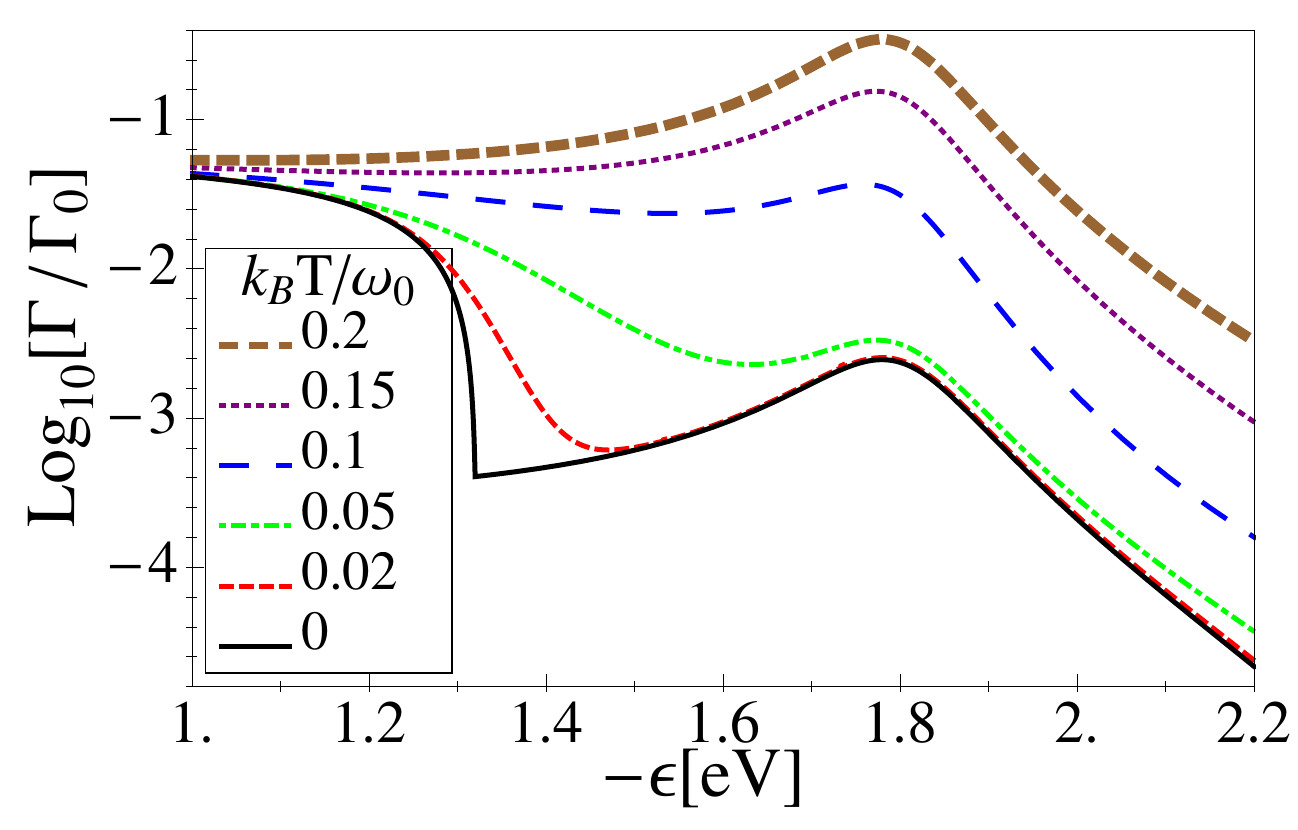}
\caption{The logarithmic total emission rate at the bias voltage $V=1.32\ \textrm{V}$ for different temperatures. The kink at the bias voltage becomes more distinct at lower temperature. 
The SPP resonance energy is taken to be $\omega_{0}=1.8$ eV, $\lambda=0.2$, and the broadening is $\eta=0.2$ eV.}
\label{fig:with temperature}
\end{figure}

We investigate the temperature dependence of the total rate in Fig.~\ref{fig:with temperature} in logarithmic scale, in which the black line shows the zero temperature case, giving the clear kink at the bias voltage $eV$, described in Ref.~\onlinecite{Fei:14}. 
%
%
%
%
%
%
%
Figure~\ref{fig:with temperature} shows how the rate sensitively depends on the temperature; the clear kink at the bias voltage is quickly softened even at small finite temperatures, and the strong effect of the temperature appears when the temperature has the same order of the factor $eV-\omega_0$, leading to the single overbias peak as the temperature is increased. 
%

Moreover, for comparison with the experimental results obtained by G. Schull and co-workers [\onlinecite{Schull:09}], 
we need first to determine the coupling parameter $\lambda$, which determines the weight between the Gaussian and non-Gaussian contributions and the width of the SPP resonance $\eta$. 
%
%
%
%
%
%
\begin{figure}[h!]
%
\includegraphics[width=0.99\columnwidth,clip=true]{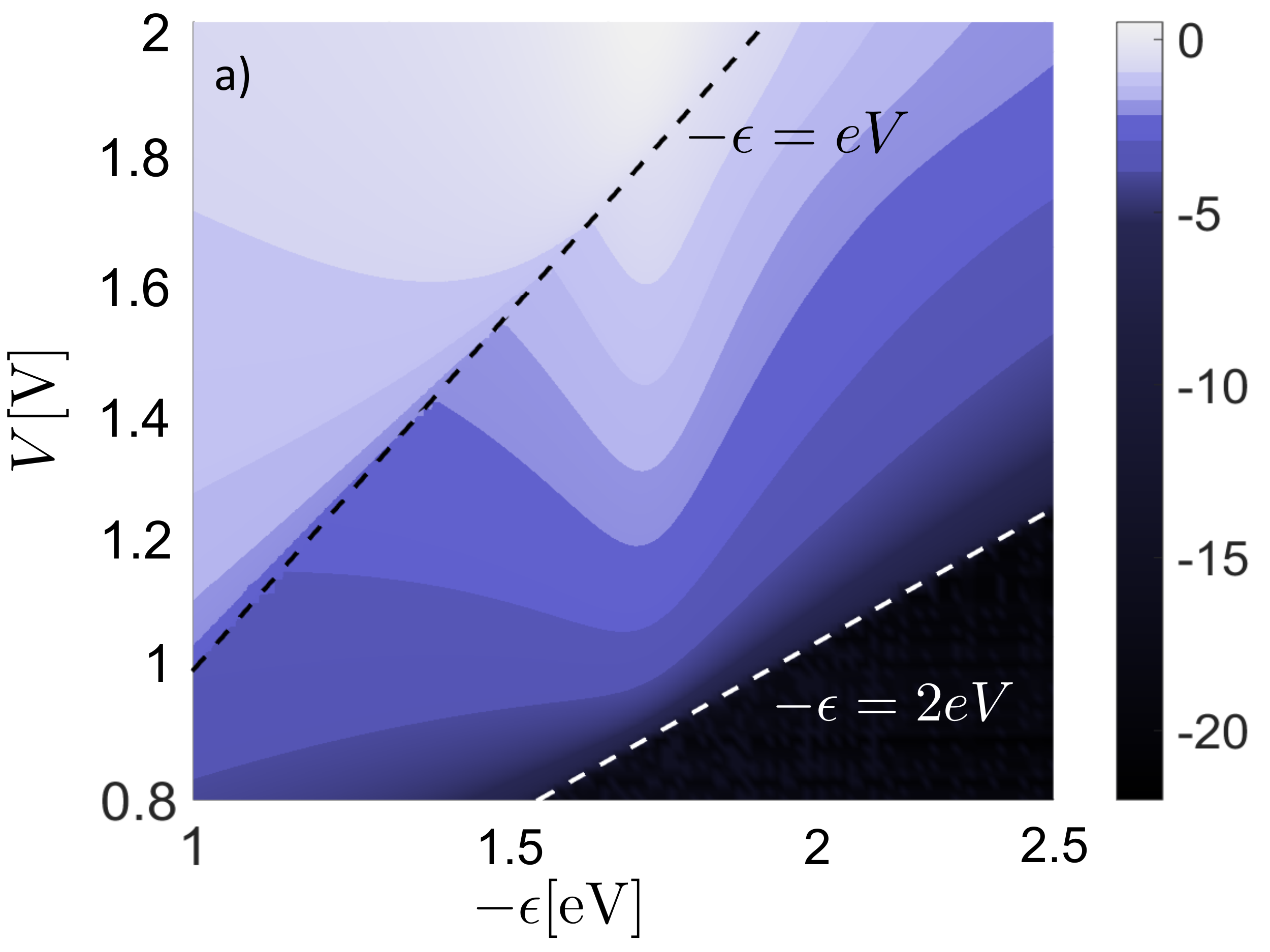} \\
\includegraphics[width=0.99\columnwidth,clip=true]{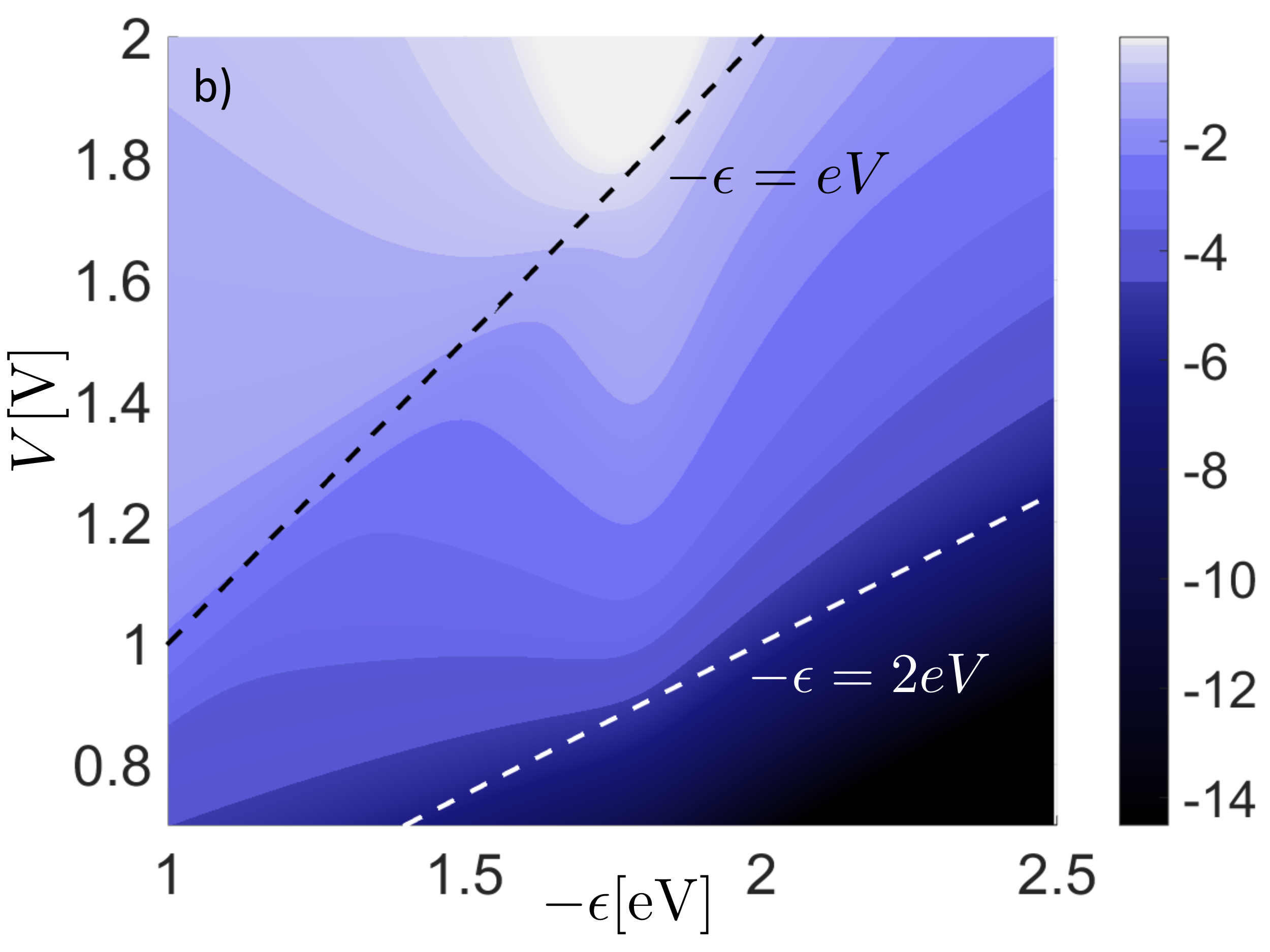}
\caption{The light emission spectrum on a logarithmic scale as a function of bias voltage at (a) the low temperature $\beta \omega_{0}=3000$ and (b) room temperature $\beta \omega_{0}=72$. In panel (a), the clear threshold behavior reproduces the experimental findings \cite{Schull:09} for the parameter $\lambda=0.227$. 
In panel (b), the threshold at the bias voltage $-\epsilon=eV$ is less evident.
This behavior depends sensitively on the temperature. Here, we use the parameter $\lambda=0.2$.
In both cases, the SPP resonance energy is taken to be $\omega_{0}=1.8$ eV, and the broadening is $\eta=0.2$ eV.}
\label{fig:fitting}
\end{figure}

The width can be directly obtained from the experimental results in Ref.~\onlinecite{Schull:09}, resulting in $\eta \approx 0.2\omega_{0}$. 
The coupling parameter is determined by scaling the peak value at $-\epsilon=\omega_{0}$ for the low bias $V=1.32\ \textrm{V}$ by a factor of 300 versus the peak at the bias $V=2.15\ \textrm{V}$, and this yields $\lambda=0.227$.
The resulting voltage- and energy-dependent emission rate is shown in Fig.~\ref{fig:fitting}(a) at the experimental temperature 
$T\simeq 7\ \textrm{K}$. For comparison, we also show the rate at room temperature $T\simeq 300\ \textrm{K}$ for $\lambda=0.2$ in Fig.~\ref{fig:fitting}(b).

%
%
Since the experimental temperature is very low compared to the frequency scale of the SPP mode $\omega_0\simeq 20.9 \,\, 10^{3}K$, 
the rate in Fig.~\ref{fig:fitting}(a) exhibits a distinct threshold at $-\epsilon=eV$, and the clear overbias peaks at the SPP resonance due to the non-Gaussian contributions, which gives a good explanation and agreement with Ref.~\onlinecite{Schull:09}. 
By contrast, at room temperature [Fig.~\ref{fig:fitting}(b)], we find that the sharp threshold behavior at $-\epsilon=eV$ has been weakened and is relaxing into the overbias SPP resonance due to the smoothed distribution function under the temperature effect. Meanwhile, the temperature effect has also sensitively hidden the two-electron energy cutoff line $-\epsilon=2eV$, leading to the long and small tail into the energy larger than $2eV$.
%


%
%
%
%
%
%
%
\begin{figure}[hbtp!]
\centering
\includegraphics[width=0.99\columnwidth,clip=true]{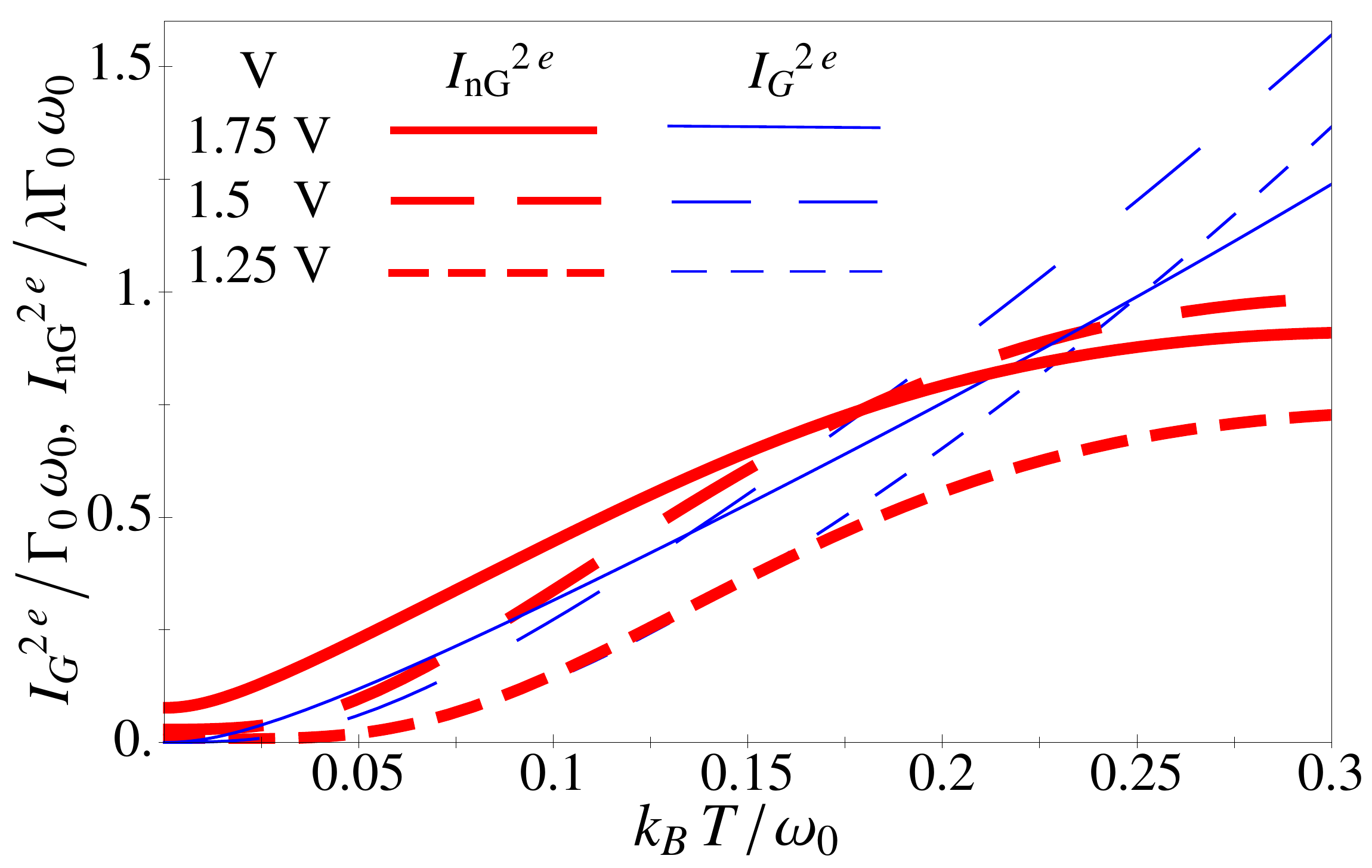}
\caption{The temperature dependence of the scaled intensity for the Gaussian and non-Gaussian contribution. The SPP resonance energy is taken to be $\omega_{0}=1.8$ eV with the broadening $\eta=0.2$ eV. }
\label{fig:Intensity}
\end{figure}

Furthermore, we study the properties of the intensity of the overbias light emission as a function of the temperature. Since the non-Gaussian part has the prefactor $\lambda=g_c z_0^{2}$ compared to the Gaussian part, we consider the Gaussian and non-Gaussian emission separately and define their intensities as $I_{\mathrm{G,nG}}^{2e}=\int ^{2eV}_{eV} \Gamma_{\mathrm{G,nG}} d\epsilon$, respectively. In Fig.~\ref{fig:Intensity}, we observe that both the Gaussian and non-Gaussian intensities increase with temperature in the temperature range shown in the figure. It is interesting to note that for high enough temperature the heating effect smears out the Fermi edge and leads to a saturation of the non-Gaussian emission. Furthermore, we find that the intensities do not increase monotonically with the bias voltages.
Hence, it would be interesting to study the temperature dependence of the overbias light emission, in order to distinguish thermally induced emission from the pure quantum effect at low temperatures. 

\section{Conclusion}

To summarize, motivated by the experimental observation of photons emitted by tunnel junctions carrying the energy larger than the bias voltage $|\epsilon|>eV$, we have developed a theoretical model to describe the electron-SPP mode interaction based on the dynamical Coulomb blockade theory.

In combination with the Keldysh path integral formalism, by treating the Gaussian and non-Gaussian contributions separately, our theory has shown that the non-Gaussian fluctuations give rise to the overbias photon emission, which can explain and reproduce the experimentally observed photon emission with energies larger than the single-particle energy limit $eV$. Furthermore, due to the smeared edge of the Fermi distribution function at finite temperature, our result also shows that the electron tunneling is sensitively affected by the temperature, thus influencing the overbias emission. The critical point at the bias voltage $-\epsilon=eV$ is strongly weakened, and the overbias peak becomes a mixture of the Gaussian and non-Gaussian noise. In addition, we also consider the interesting case when the bias voltage is far from the SPP resonance; here we set the resonance close to the two-electron energy limit, and we argue that this regime is suitable to distinguish the Gaussian and non-Gaussian contributions even at finite temperature and in the case of sharp resonance. Finally, we investigate the temperature dependence of the photon intensities in the overbias region at different bias voltages and show that it allows us to distinguish the quantum emission from a pure heating effect.  
%

In conclusion, our work enables us to model the light emission due to the electron-SPP mode interaction in nanosize contacts 
and can be applied to more complex junctions.\\

\acknowledgments

This work was supported by the DFG through SFB 767, the Center of Applied Photonics (CAP), the Kurt Lion Foundation, and the Zukunftskolleg of the University of Konstanz.

\appendix

\section{Expansion}\label{sec:appendix}

Here, we describe some useful intermediate results for the derivation of the non-Gaussian rate, Eq.~(\ref{eq:nG-rate}), and the expansion of the action of the tunnel conductor $\mathbb{S}_{c}$ to the fourth order in the fluctuating fields.
According to the Gaussian averages $\langle \langle \Phi_{\omega} \rangle \rangle$ and $\langle \langle \Phi_{\omega}\Phi_{-\omega}^{T} \rangle \rangle$, we obtain
\begin{eqnarray*}
\langle \langle \varphi^{+}(\omega) \rangle \rangle &=& i\alpha[Y(\omega)e^{i\omega t}-X(\omega)], \\
\langle \langle \varphi^{-}(\omega) \rangle \rangle &=& i\alpha[Q(\omega)e^{i\omega t}-P(\omega)], \\
\langle \langle \varphi^{+}_{\omega} \varphi^{+}_{\omega'} \rangle \rangle &=& 2\pi X(\omega) \delta(\omega+\omega'), \\
\langle \langle \varphi^{+}_{\omega} \varphi^{-}_{\omega'} \rangle \rangle &=& 2 \pi Y(\omega) \delta(\omega+\omega'), \\
\langle \langle \varphi^{-}_{\omega} \varphi^{+}_{\omega'} \rangle \rangle &=& 2 \pi P(\omega) \delta(\omega+\omega'), \\
\langle \langle \varphi^{-}_{\omega} \varphi^{-}_{\omega'} \rangle \rangle &=& 2\pi Q(\omega) \delta(\omega+\omega').
\end{eqnarray*}
with
\begin{eqnarray*}
X(\omega)&=& S_{nc}(\omega)\frac{\lvert \tilde{z}_{\omega}\rvert^{2}}{\omega^{2}}+\frac{1}{\omega^2} [W(\omega) Re\{ \tilde{z}_{\omega}\}+i\omega Im\{ \tilde{z}_{\omega}\}], \\
Y(\omega)&=&  S_{nc}(\omega)\frac{\lvert \tilde{z}_{\omega}\rvert^{2}}{\omega^{2}}+\frac{1}{\omega^2} [W(\omega)-\omega] Re\{ \tilde{z}_{\omega}\}, \\ 
P(\omega)&=& S_{nc}(\omega)\frac{\lvert \tilde{z}_{\omega}\rvert^{2}}{\omega^{2}}+\frac{1}{\omega^2} [W(\omega)+\omega] Re\{ \tilde{z}_{\omega}\}, \\ 
Q(\omega)&=& S_{nc}(\omega)\frac{\lvert \tilde{z}_{\omega}\rvert^{2}}{\omega^{2}}+\frac{1}{\omega^2} [W(\omega) Re\{ \tilde{z}_{\omega}\}-i\omega Im\{ \tilde{z}_{\omega}\}], 
\end{eqnarray*}
with $S_{nc}(\omega)=g_{c}[\frac{1}{2}W(\omega+eV)+\frac{1}{2}W(\omega-eV)-W(\omega)]$.

After performing the symmetrization over $\omega$, we obtain
\begin{widetext}
\begin{displaymath}
\begin{aligned}
&\langle \langle \mathbb{S}_{c}^{(4)} \rangle \rangle=- \frac{i \pi \alpha^2 g_{c} }{16}\iint d \omega d \omega' \Bigg\{\![Y(\omega')e^{i\omega' t}\!-\!X(\omega')][Y(\!-\omega')e^{-i\omega' t}\!-\!X(\!-\omega')]\!\Big[\! X(\omega) [-\!F(0)\!+\!2F^{s}(\omega)\!+\!2F^{s}(\omega')\!-\!2F^{ss}(\!-\omega\!-\!\omega')] \\
&\quad \quad+Q(\omega)F(0)-P(\omega)F(\omega)-Y(\omega)F(-\omega) \Big] \\
&+[Q(\omega')e^{i\omega' t}-P(\omega')][Q(-\omega')e^{-i\omega' t}-P(-\omega')]\Big[Q(\omega)[-\!F(0)\!+\!2F^{s}(\omega)\!+\!2F^{s}(\omega')\!-\!2F^{ss}(\!-\omega\!-\!\omega')] \\
&\quad \quad-P(\omega)F(\omega)-Y(\omega)F(-\omega)+X(\omega)F(0)\Big] \\
&+[Y(\omega')e^{i\omega' t}-X(\omega')][Q(-\omega')e^{-i\omega' t}-P(-\omega')]\Big[-[Q(\omega)+X(\omega)] F(-\omega')+Y(\omega)F(-\omega-\omega')+P(\omega)F(\omega-\omega')\Big] \\
& + [Q(\omega')e^{i\omega' t}-P(\omega')][Y(-\omega')e^{-i\omega' t}-X(-\omega')]\Big[ -[Q(\omega)+X(\omega)]F(\omega')+Y(\omega)F(-\omega+\omega')+P(\omega)F(\omega+\omega')\Big] \Bigg\}
\end{aligned}
\end{displaymath}
\end{widetext}
with the defined functions $F^{s}(\omega)=[F(\omega)+F(-\omega)]/2$ and $F^{ss}(-\omega-\omega')=[F(-\omega-\omega')+F(-\omega+\omega')+F(\omega-\omega')+F(\omega+\omega')]/4$, in which $F(\omega)=F_{1}(\omega)+F_{2}(-\omega)=(-\omega-eV)+W(-\omega-eV)+(-\omega+eV)+W(-\omega+eV)$ as given in the text.

One can show that the terms proportional to $e^{i\omega' t}$ and the ones proportional to $e^{-i\omega' t}$, are interchanged under the operation $\omega' \rightarrow -\omega'$. Using $\int e^{i\omega t} e ^{i \epsilon t}dt=2 \pi \delta(\omega+\epsilon)$ and keeping the terms in the lowest order of $g_{c}Z_{0}^2$, the non-Gaussian rate Eq.~(\ref{eq:nG-rate}) can be expressed as
\begin{widetext}
\begin{displaymath}
\begin{aligned}
\Gamma_{\mathrm{nG}}^{(4)}=&\frac{\pi^2\alpha^2 |\mathcal{T}|^2 g_{c}}{4}\int \! d \omega \Bigg\{ Y(-\epsilon)X(\epsilon) \Big[ X(\omega)[-\!F(0)\!+\!2F^{s}(\omega)\!+\!2F^{s}(\epsilon)\!-\!2F^{ss}(-\omega\!+\!\epsilon)]\!-P(\omega)F(\omega)\!-Y(\omega)F(\!-\omega)\!+Q(\omega)F(0) \Big] \\
&+Q(-\epsilon)P(\epsilon) \Big[ Q(\omega) [-\!F(0)\!+\!2F^{s}(\omega)\!+\!2F^{s}(\epsilon)\!-\!2F^{ss}(-\omega\!+\!\epsilon)]\!-P(\omega)F(\omega)\!-Y(\omega)F(\!-\omega)\!+X(\omega)F(0)\Big] \\
&+Y(-\epsilon)P(\epsilon) \Big( -[Q(\omega)+X(\omega)]F(\epsilon)+Y(\omega)F(-\omega+\epsilon)+P(\omega)F(\omega+\epsilon)\Big) \\
&+Q(-\epsilon)X(\epsilon) \Big( -[Q(\omega)+X(\omega)]F(-\epsilon)+Y(\omega) F(-\omega-\epsilon)+P(\omega)F(\omega-\epsilon)\Big) \Bigg\}.
\end{aligned}
\end{displaymath}
\end{widetext}
This expression can be cast as Eq.~(\ref{eq:nG-rate}) in the main text after replacing all the functions, i.e., $X, Y, P, Q$ and $F$, by their definitions.




\begin{thebibliography}{9}
\bibitem{notes}
Notice that, in the small tunneling limit $g_c \ll 1$ here considered, we have the Fano factor $F=1$ for the current noise.
\end{thebibliography}
\end{document}